\documentclass[a4paper,12pt]{article}
\pdfoutput=1 
\usepackage{amsfonts,amsmath,amsthm,amssymb,amscd}
\usepackage{natbib}
\usepackage[nobottomtitles*]{titlesec}
\titleformat{\section}{\large\sf}{\thesection}{1em}{}
\titleformat{\subsection}{\sf}{\thesubsection}{1em}{}
\titlespacing*{\section}{0pt}{*3.5}{1.5ex}[\fill]
\titlespacing*{\subsection}{0pt}{*2}{0ex}[\fill]

\usepackage{needspace}
\usepackage{newpxtext,newpxmath}
\usepackage{noto}
\usepackage{microtype}
\usepackage{xcolor}
\usepackage{tikz}
\usepackage{pgfplots}
\usetikzlibrary{shapes} 
\pgfplotsset{compat=1.16}
\usetikzlibrary{arrows.meta}
\usepackage{graphicx}
\usepackage[margin=1.8em,font={sf,footnotesize},justification=justified,labelfont=bf,labelsep=quad]{caption}
\usepackage[capbesideposition={right,bottom},capbesidesep=qquad]{floatrow}
\def\FCM#1{{\small$#1$}} 
\usepackage{wrapfig} 
\usepackage[colorlinks,linkcolor=blue,citecolor=purple]{hyperref}
\newcommand{\reals}{\mathbb{R}}

\newcommand{\eps}{\varepsilon}
\newcommand{\T}{^{\kern-.04em\top\kern-.1em}}
\newcommand{\1}{\normalfont\textup{\textbf1}}  
\newcommand{\0}{\normalfont\textup{\textbf0}} 
\newcommand{\conv}{\textup{\small\textsf{conv\,}}}

\newcommand{\supp}{\textup{\small\textsf{supp}}}
\newcommand{\br}{\textup{\small\textsf{bestresp}}}
\newcommand{\sign}{\textup{\small\textsf{sign}}}
\def\ci#1{{\lower3.4pt\hbox{%
\begin{tikzpicture}[scale=1]
\draw [] (0,0) node[inner sep=1.5pt,thin,draw,circle] {\footnotesize\sf#1};
\end{tikzpicture}}}}

\newcommand\commentout[1]{}
\newtheorem{theorem}{Theorem}[]

\newtheorem{lemma}[theorem]{Lemma}
\newtheorem{proposition}[theorem]{Proposition}

\theoremstyle{definition}
\newtheorem{definition}[theorem]{Definition}

\newtheorem{algorithm}[theorem]{Algorithm}

\DeclareFloatVCode{afterfloat}{\vspace{-10pt}}
\newdimen\einr\einr1.8em
\newdimen\eeinr\eeinr3.8em
\newdimen\rmeinr\rmeinr1.8em
\newdimen\tmp 
\newcommand{\abs}[1]{\par\hangafter=1\hangindent=\einr
  \noindent\hbox to\einr{#1\hfill}\ignorespaces}
\def\aabs#1{\par\hangafter=1\hangindent=\eeinr
    \noindent\hbox to\eeinr{\strut\hskip\einr#1\hfill}\ignorespaces}

\newcommand\bullitem{\tmp\einr\einr\rmeinr\abs{\raise.17ex\hbox{\kern7pt\scriptsize$\bullet$}}\einr\tmp}
\parindent\einr
\def\proof{\noindent{\textit{Proof.\enspace}}}

\def\endproof{\hfill\strut\nobreak\hfill\tombstone\par\medbreak}
\def\tombstone{\hbox{\lower.4pt\vbox{\hrule\hbox{\vrule
  \kern7.6pt\vrule height7.6pt}\hrule}\kern.5pt}}

\newsavebox{\figA} 
\newsavebox{\figB} 
\newsavebox{\figC} 
\title{Finding Nash Equilibria of Two-Player Games}
\author{Bernhard von Stengel}
\date{\normalsize 
Department of Mathematics, London School of Economics,
London WC2A 2AE,
United Kingdom. Email: \textsf{b.von-stengel@lse.ac.uk}
\\[2ex]
February 9, 2021} 
\begin{document} 
\maketitle

\begin{abstract}
This paper is an exposition of algorithms for finding one or
all equilibria of a bimatrix game (a two-player game in
strategic form) in the style of a chapter in a graduate
textbook.
Using labeled ``best-response polytopes'', we present the
Lemke-Howson algorithm that finds one equilibrium.
We show that the path followed by this algorithm has a
direction, and that the endpoints of the path have opposite
\textit{index}, in a canonical way using determinants.
For reference, we prove that a number of notions of
\textit{nondegeneracy} of a bimatrix game are equivalent.
The computation of all equilibria of a general bimatrix
game, via a description of the maximal Nash subsets of the
game, is canonically described using ``complementary pairs''
of faces of the best-response polytopes. 
\end{abstract}

\section{Introduction}

\noindent
A \textit{bimatrix game} is a two-player game in strategic form,
specified by the two matrices of payoffs to the row player
and column player.
This article describes algorithms that \textit{find} one or
all Nash equilibria of such a game.

The game gives rise to two suitably \textit{labeled
polytopes} (described in Section~\ref{s-poly}), which help
finding its Nash equilibria.
This geometric structure is also very informative and
accessible, for example for the construction of $3\times 3$
games with a certain equilibrium structure, which is much
more varied than for $2\times 2$ games.
It also provides an elementary and constructive proof for
the existence of a Nash equilibrium for a bimatrix game
via the algorithm by \citet{LH}.
In Section~\ref{s-lh}, we first explain this algorithm
following the exposition by \citet{Shapley1974}, in
particular using the subdivision of the mixed strategy
simplices $X$ and $Y$ into best-response regions,
and the construction of
$\tilde X$ and $\tilde Y$ in Section~\ref{s-lh} and
Figure~\ref{flh}, which extends $X\times Y$ with an
``artificial equilibrium''.
In Section~\ref{s-LHpoly}, we then give a more concise
description using polytopes.
Section~\ref{s-index} gives a canonical proof that the
endpoints of Lemke-Howson paths have opposite
\textit{index}.
The index is here defined in an elementary way using
determinants (Definition~\ref{d-index}).
In Section~\ref{s-nondegen}, we show that a number of known
definitions of \textit{nondegeneracy} of a bimatrix game are in
fact equivalent.
Section~\ref{s-pivot} shows how to implement the
Lemke-Howson algorithm by ``complementary pivoting'', even
when the game is degenerate.
Section~\ref{s-maxnash} describes the structure of Nash
equilibria of a general bimatrix game.

An undergraduate text, in even more detailed style and
avoiding advanced mathematical machinery, is \citet{vS2021}.
This article continues Chapter~9 of that book, with the
proof of the direction of a Lemke--Howson path and the
concept of the index of an equilibrium, and the detailed
discussion of nondegeneracy.
Earlier expositions of this topic are \citet{vS2002}, which
gives additional historical references, and \citet{vS2007}.
Compared to these surveys, the following expository results
are new:

\bullitem
The definition of the index of a Nash
equilibrium in a nondegenerate game, and the very canonical
proof that opposite endpoints of Lemke-Howson paths have
opposite index in Theorem~\ref{t-opp}.
Essentially, this is a much more accessible version of the
argument by \citet{Shapley1974}.

\bullitem
The equivalent definitions of nondegeneracy in
Theorem~\ref{t-eq}. 

\bullitem
A cleaner presentation of maximal Nash subsets, adapted from
\citet{ARSvS}, in Proposition~\ref{p-faces}.
\einr\rmeinr

\section{Bimatrix games and the best response condition}
\label{s-bimatrix}

We use the following notation throughout.
Let $(A,B)$ be an $m\times n$ bimatrix game, that is, $A$
and $B$ are $m\times n$ matrices of payoffs to the row
player~1 and column player~2, respectively.
This is a two-player game in strategic form (also called
``normal form''), which is played by a simultaneous choice
of a row $i$ by player~1 and column~$j$ by player~2, who
then receive the entries $a_{ij}$ of the matrix $A$, and
$b_{ij}$ of $B$, as respective payoffs.
The payoffs represent risk-neutral utilities, so when facing
a probability distribution, the players want to maximize
their expected payoff.
These preferences do not depend on positive-affine
transformations, so that $A$ and $B$ can be assumed to have
nonnegative entries. 
In addition, as inputs to an algorithm they are assumed to
be rationals or just integers. 

All vectors are column vectors, so an $m$-vector vector $x$
(that is, an element of $\reals^m$) is treated as an
$m\times 1$ matrix, with components $x_1,\ldots,x_m$.
A~scalar is treated as a $1\times 1$ matrix, and therefore
multiplied to the right of a column vector and to the left
of a row vector.
A~\textit{mixed strategy} $x$ for player~1 
is a probability distribution on the rows of the game,
written as an $m$-vector of probabilities.
Similarly, a mixed strategy $y$ for player~2 is an
$n$-vector of probabilities for playing the columns of the
game.
Let $\0$ be the all-zero vector and let $\1$ be the all-one
vector of appropriate dimension.
The transpose of any matrix $C$ is denoted by $C\T$, so
$\1\T$ is the all-one row vector.
Inequalities like $x\ge\bf 0$ between two vectors hold for
all components.
Let $X$ and $Y$ be the mixed-strategy sets of the two
players,
\begin{equation}
\label{XY}
X = \{x\in\reals^m\mid x\ge\0,~\1\T x=1\,\},
\qquad
Y = \{y\in\reals^n\mid y\ge\0,~\1\T y=1\,\}.
\end{equation}
The \textit{support} $\supp(z)$ of a mixed strategy $z$ is
the set of pure strategies that have positive probability,
so $\supp(z)=\{\,k\mid z_k>0\}$.

A \textit{best response} to the mixed strategy $y$
of player~2 is a mixed strategy $x$ of player~1 that maximizes 
his expected payoff $x^{\top} A y$.
Similarly, a best response $y$ of player~2 to $x$  
maximizes her expected payoff $x^{\top} B y$.
A \textit{Nash equilibrium} or just \textit{equilibrium} is
a pair $(x,y)$ of mixed strategies that are best responses
to each other.
The following 
proposition states that a mixed strategy $x$ is a best
response to an opponent strategy~$y$ if and only if all pure
strategies in its support are pure best responses to~$y$.
The same holds with the roles of the players exchanged.

\begin{proposition}[Best response condition]
\label{p-bestresponse}
Let $x$ and $y$ be mixed strat\-e\-gies of player $1$ and~$2$,
respectively.
Then $x$ is a best response to $y$ if and only if for all
$i=1,\ldots,m$,
\begin{equation}
\label{bestresp}
x_i>0~~\Rightarrow~~ (Ay)_i=u=\max\{\,(Ay)_k\mid
k=1,\ldots,m\}.
\end{equation}
\end{proposition}

\proof
$(Ay)_i$ is the $i$th component of $Ay$, which is
the expected payoff to player~1 when playing row~$i$.
Then
\[
x^{\top} Ay=\sum_{i=1}^m x_i\, (Ay)_i
= \sum_{i=1}^m x_i\, (u-(u-(Ay)_i)
= u - \sum_{i=1}^m x_i\, (u-(Ay)_i).
\]
So $x^{\top} Ay\le u$ because $x_i\ge 0$ and 
$u-(Ay)_i\ge 0$ for all $i=1,\ldots,m$,
and $x^{\top} Ay=u$ if
and only if $x_i>0$ implies $(Ay)_i=u$, as claimed.  
\endproof

Proposition~\ref{p-bestresponse} is useful in a number of
respects.
First, by definition, $x$ is a best response to $y$ if and
only if $x\T Ay\ge \hat x\T Ay$ for all other mixed
strategies $\hat x$ in~$X$ of player~1, where $X$ is an
infinite set.
In contrast, (\ref{bestresp}) is a finite condition, which
only concerns the pure strategies $i$ of player~1, which
have to give maximum payoff $(Ay)_i$ whenever $x_i>0$.
For example, in the $3\times 2$ game
\begin{equation}
\label{example}
A=\left[\,\begin{matrix}3 & 3 \cr 2&5\cr
0&6\cr\end{matrix}\,\right],\qquad
B=\left[\,\begin{matrix}3&2\cr 2&4\cr 3&
0\cr\end{matrix}\,\right],
\end{equation} 
if $y=(\frac 13,\frac23)\T$, then $Ay=(3,4,4)\T$.
(From now on, we omit for brevity the transposition when writing down
specific vectors, as in $y=(\frac 13,\frac23)$.)
Then player~1's pure best responses against $y$ are the
second and third row, and $x=(x_1,x_2,x_3)$ in~$X$ is a
best response to $y$ if and only if $x_1=0$.
In order for $x$ to be a best response against $y$, the pure
best responses $2$ and $3$ can be played with
\textit{arbitrary probabilities} $x_2$ and $x_3$.
(As part of an equilibrium these probabilities will here be
unique in order to ensure the best response condition for
the \textit{other} player.)
Second, as the proof of Proposition~\ref{p-bestresponse}
shows, mixing cannot improve the payoff of a player (here of
player~1), which is just a ``weighted average'' of the
expected payoffs $(Ay)_i$ with the weights $x_i$ for the
rows~$i$.
This payoff is maximal only if only the maximum
pure-strategy payoffs $(Ay)_i$ have positive weight.

We denote by $\br(z)$ the set of \textit{pure best
responses} of a player against a mixed strategy $z$ of the
other player, so
$\br(y)\subseteq\{1,\ldots,m\}$ if $y\in Y$ and
$\br(x)\subseteq\{1,\ldots,n\}$ if $x\in X$.
Then (\ref{bestresp}) states that $x$ is a best response
to~$y$ if and only if 
\begin{equation}
\label{brsub}
\supp(x)\subseteq\br(y).
\end{equation}
This condition applies also to games with any finite number
of players:
If $s$ is any one of $N$ players who plays the mixed
strategy $x^s$, with the tuple of the ${N-1}$ mixed strategies
of the remaining players denoted by $x^{-s}$, then $x^s$ is
a best response against $x^{-s}$ if and only if 
\begin{equation}
\label{brsubN}
\supp(x^s)\subseteq\br(x^{-s}).
\end{equation}
The proof of Proposition~\ref{p-bestresponse} still applies,
where instead of $(Ay)_i$ for a pure strategy~$i$ of
player~$s$ one has to use the expected payoff to player~$s$
when he uses strategy~$i$ against the tuple $x^{-s}$ of
mixed strategies of the other players.
For more than two players, $N>2$, that expected payoff
involves products of the mixed strategy probabilities $x^r$
for the other players $r$ in $N-\{s\}$ and is therefore
non\-linear. 
The resulting polynomial equations and inequalities make the
structure and computation of Nash equilibria for such games
much more complicated than for two players, where the
expected payoffs $(Ay)_i$ are linear in the opponent's mixed
strategy~$y$.
We consider only two-player games here.

Proposition~\ref{p-bestresponse} is used in algorithms that
find Nash equilibria of the game.
One such approach is to consider the different possible
\textit{supports} of mixed strategies.
All pure strategies in the support must have maximum, and
hence equal, expected payoff to that player.
This leads to equations for the probabilities of the
\textit{opponent's} mixed strategy.
In the above example (\ref{example}), the mixed strategy
$y=(\frac 13,\frac23)$ has any $x=(0,x_2,x_3)\ge\0$ with
$x_2+x_3=1$ as a best response.
In order for $y$ to be a best response against such an $x$,
the two columns have to have maximal and hence equal payoff
to player~2, that is, $2x_2+3x_3=4x_2+0x_3$,
which has the unique solution $x_2=\frac 35$, $x_3=\frac 25$
and expected payoff $\frac{12}5$ to player~2.
Hence, $(x,y)$ is an
equilibrium, which we denote for later reference by $(a,c)$,
\begin{equation}
\label{ac}
\textstyle
(a,c)=((0,\frac35,\frac25),(\frac13,\frac23)).
\end{equation}
Here the mixed strategy $y$ of player~2 is uniquely
determined by the condition $2y_1+5y_2=0y_1+6y_2$ that the
two bottom rows give equal expected payoff to player~1.

A second mixed equilibrium $(x,y)$ is given if the support
of player~1's strategy consists of the first two rows, which
gives the equation $3y_1+3y_2=2y_1+5y_2$ with the unique
solution
$y=(\frac23,\frac13)$ and thus $Ay=(3,3,2)$.
With $x=(x_1,x_2,0)$ the equal payoffs to player~2 for her 
two columns give the equation $3x_1+2x_2=2x_1+4x_2$ with
unique solution $x=(\frac23,\frac13,0)$.
Then $(x,y)$ is an equilibrium, for later reference denoted
by $(b,d)$,
\begin{equation}
\label{bd}
\textstyle
(b,d)=((\frac23,\frac13,0),(\frac23,\frac13)).
\end{equation}
A third, pure-strategy Nash equilibrium of the game is
$((1,0,0),(1,0))$.

The support set $\{1,3\}$ for the mixed strategy of player~1
does not lead to an equilibrium, for two reasons.
First, player~2 would have to play $y=(\frac12,\frac12)$
to make player~1 indifferent between row~1 and row~3.  
But then the vector of expected payoffs to player~1 is
$Ay=(3,\frac72,3)$, so that rows 1 and~3 give the same
payoff to player~1 but not the maximum payoff for all rows.
Second, player~2 needs to be indifferent between her two 
strategies (because player~1's best response to a pure
strategy is unique and cannot have the support $\{1,3\}$).
The corresponding equation $3x_1+3x_3=2x_1$ (together with
$x_1+x_3=1$) has the solution $x_1=\frac32$, $x_3=-\frac12$,
so $x$ is not a vector of probabilities.  

In this ``support testing'' method, it normally suffices 
to consider supports of equal size for the two players.
For example, in (\ref{example}) it is not necessary to
consider a mixed strategy $x$ of player~1 where all three
pure strategies have positive probability, because player~1
would then have to be indifferent between all these.
However, a mixed strategy $y$ of player~1 is already
uniquely determined by equalizing the expected payoffs
for two rows, and then the payoff for the remaining row is
already different.
This is the typical, ``nondegenerate'' case, according to
the following definition.

\begin{definition}
\label{d-nondegenerate}
A two-player game is called {\em nondegenerate} if no
mixed strategy~$z$ of either player of support size~$k$ has
more than $k$ pure best responses, that is,
$|\br(z)|\le|\supp(z)|$. 
\end{definition}

In a \textit{degenerate} game, Definition~\ref{d-nondegenerate}
is violated, for example if there is a pure strategy that
has two pure best responses.
For the moment, we only consider nondegenerate games,
where the players' equilibrium strategies have equal sized
support, which is immediate from Proposition~\ref{p-bestresponse}:

\begin{proposition}
In any Nash equilibrium $(x,y)$ of a nondegenerate
bimatrix game, $x$ and $y$ have supports of equal size.
\end{proposition}

\proof
Condition (\ref{brsub}), and the analogous condition
$\supp(y)\subseteq\br(x)$, give 
\[
|\supp(x)|\le|\br(y)|\le |\supp(y)|\le|\br(x)|\le |\supp(x)|
\] 
so we have equality throughout.
\endproof

The ``support testing'' algorithm for finding equilibria
of a nondegenerate bimatrix game considers any two 
equal-sized supports of a potential equilibrium, equalizes
their payoffs $u$ and $v$, and then checks whether $x$ and $y$
are mixed strategies and $u$ and $v$ are maximal payoffs.

\begin{algorithm}[Equilibria by support enumeration]
\label{a-suppenum}
\ 
{\em Input:}
An $m\times n$ bimatrix game $(A,B)$ that is nondegenerate.
{\em Output:} All Nash equilibria of the game.
{\em Method:} 
For each $k=1,\ldots,\min\{m,n\}$ and each pair $(I,J)$ of
$k$-sized sets of pure strategies for the two players, solve
(with unknowns $x,v,y,u$)
the equations
$\sum_{i\in I}x_ib_{ij}=v$ for $j\in J$, $\sum_{i\in I}x_i=1$,
$\sum_{j\in J}a_{ij}y_j=u$ for $i\in I$, $\sum_{j\in J}y_j=1$,
and subsequently check that $x\ge\hbox{\bf0}$, $y\ge\hbox{\bf0}$,
and that (\ref{bestresp}) holds for $x$ and analogously~$y$.
If so, output $(x,y)$.
\end{algorithm}

The linear equations considered in this algorithm may not
have solutions, which then mean no equilibrium for that
support pair.
Nonunique solutions can occur for degenerate games, which
have underdetermined systems of linear equations for
equalizing the opponent's expected payoffs (see
Theorem~\ref{t-eq}(f) below).

\section{Equilibria via labeled polytopes}
\label{s-poly}

Algorithm~\ref{a-suppenum} can be improved because equal
payoffs for the pure strategies in a potential equilibrium
support do not imply that these payoffs are also optimal,
for example against the mixed strategy $y=(\frac12,\frac12)$
in example (\ref{example}).
By using suitable linear inequalities, one can capture this
additional condition automatically.
This gives rise to ``best-response polyhedra'', which have
equivalent descriptions via ``best-response regions'' and
``best-response polytopes''.

In this geometric approach, mixed strategies $x$ and $y$ are
considered as points in the respective mixed strategy
``simplex'' $X$ or $Y$ in (\ref{XY}).
We use the following notions from convex geometry.
An \textit{affine combination} of points $z^1,\ldots, z^k$ 
in some Euclidean space is of the form $\sum_{i=1}^k z^i\lambda_i$
where $\lambda_1,\ldots,\lambda_k$ are reals with
${\sum_{i=1}^k \lambda_i=1}$.
It is called a \textit{convex combination} if $\lambda_i\ge0$ for all~$i$.
A set of points is \textit{convex} if it is closed
under forming convex combinations.
The \textit{convex hull} of a set of points is the smallest
convex set that contains all these points.
Given points are \textit{affinely independent} if none of these
points is an affine combination of the others.
A convex set has \textit{dimension}~$d$ if and only if
it has $d+1$, but no more, affinely independent points.
A \textit{simplex} is the convex hull of a set of affinely
independent points.
The $k$th \textit{unit vector} has its $k$th component equal
to one and all other components equal to zero.
The mixed strategy simplex $X$ of player~1 in~(\ref{XY}) is
the convex hull of the $m$ unit vectors in $\reals^m$ (and
has dimension $m-1$), and $Y$ is the convex hull of the $n$
unit vectors in $\reals^n$ (and has dimension $n-1$).

\begin{figure}[hbt]
\[
\includegraphics[width=70mm]{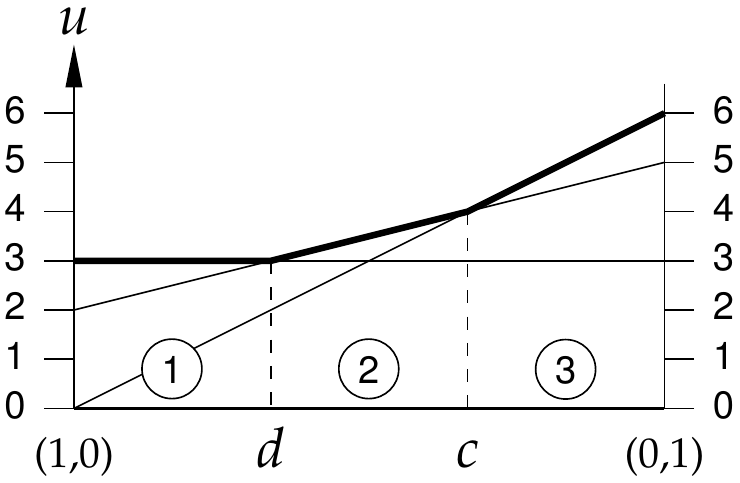}
\]
\caption{Upper envelope of expected payoffs to player~1, as
a function of the mixed strategy \FCM{y} of player~2, for
the game (\ref{example}).}
\label{f3br}
\end{figure}

For the $3\times 2$ game in (\ref{example}), $Y$ is the
line segment that connects the unit vectors $(1,0)$ and
$(0,1)$, whose convex combinations $(y_1,y_2)$ are the mixed
strategies of player~2.
The resulting expected payoffs to player~1 for his three
pure strategies are given by $3y_1+3y_2$, $2y_1+5y_2$, and
$0y_1+6y_2$.
The maximum of these three linear expressions in $(y_1,y_2)$
defines the \textit{upper envelope} of player~1's expected
payoffs, shown in bold in Figure~\ref{f3br}.
This picture shows that
row 1 is a best response if $y_2\in[0,\frac 13]$,
row 2 is a best response if $y_2\in[\frac13,\frac 23]$,
and
row 3 is a best response if $y_2\in[\frac23,1]$.
The sets of mixed strategies $y$ corresponding to these
three intervals are \textit{labeled} with the pure
strategies $1,2,3$ of player~1, shown as circled numbers in
the picture.
The point $d=(\frac23,\frac13)$ has two labels 1 and 2,
which are the two pure responses of player~1.
Similarly, point $c=(\frac13,\frac23)$ has the two labels 2
and 3 as best responses.
The picture shows also that for $y=(\frac12,\frac12)$ the
two pure strategies 1 and 3 have equal expected payoff, but
the label of this point $y$ is 2 because its (unique) best
response, row~2, has higher payoff.

\begin{figure}[hbt]
\[
\includegraphics[width=89mm]{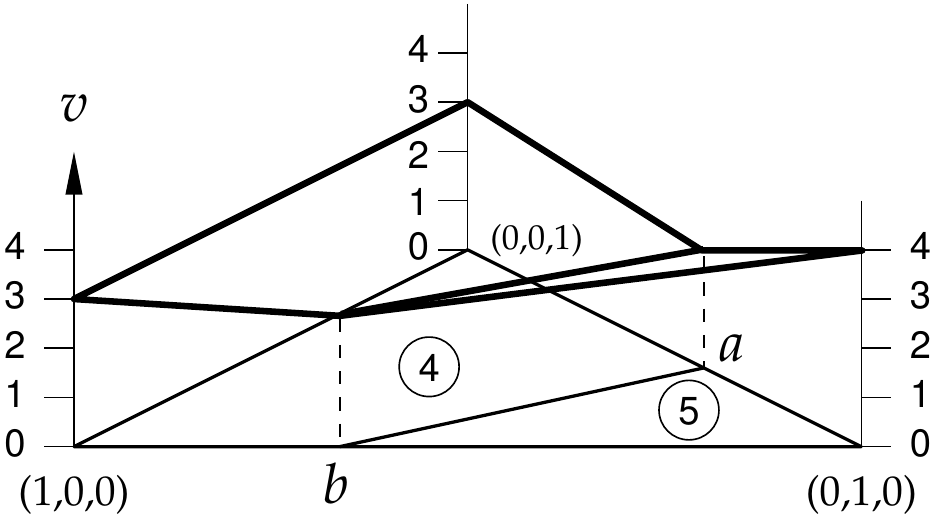}
\]
\caption{Perspective drawing of the upper envelope of
expected payoffs to player~2, as a function of the mixed
strategy \FCM{x} of player~1, for the game (\ref{example}).}
\label{f2br}
\end{figure}

We \textit{label} the pure strategies of the two players
\textit{uniquely} by giving label~$i$ to each row
$i=1,\ldots,m$, and label~$m+j$ to each column
$j=1,\ldots,n$.
In our $3\times 2$ example, the pure strategies of player~2
have therefore labels 4 and~5.
Figure \ref{f2br} shows the upper envelope for the two
strategies of player~2 for the possible mixed strategies
$x\in X$ of player~1; note that $X$ is a triangle.
As found earlier in (\ref{ac}) and (\ref{bd}), for the
points $a=(0,\frac35,\frac25)$ and $b=(\frac23,\frac13,0)$
in $X$ both columns have equal expected payoffs to player~2.
This is also the case for any convex combination of $a$
and~$b$, that is, any point on the line segment that
connects $a$ and~$b$.
This line segment is common to the two best-response regions
that otherwise partition $X$, namely the best-response
region for the first column (with label~4) that is the
convex hull of the points $(1,0,0)$, $b$, $a$, and
$(0,0,1)$, and the best-response region for the second
column (with label~5) which is the convex hull of the points
$(0,1,0)$, $a$, and $b$.
Both regions are shown in Figure~\ref{f2br}. 

\begin{figure}[hbt]
\[
\includegraphics[width=150mm]{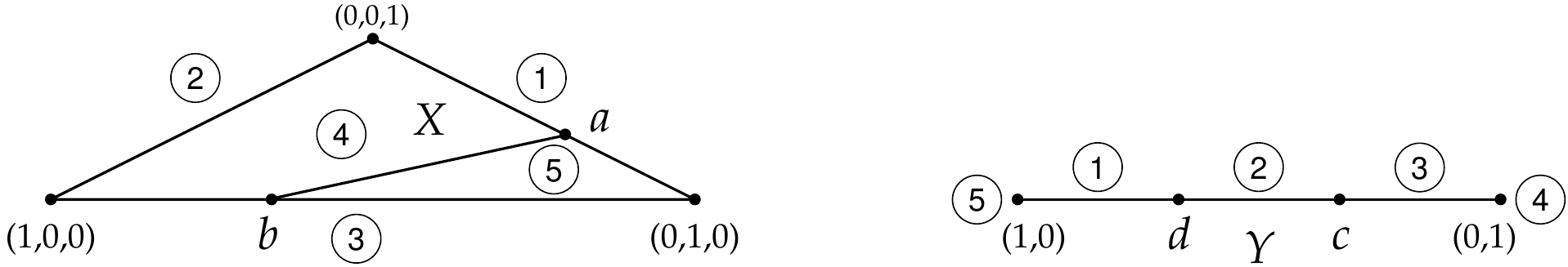}
\]
\caption{The mixed strategy sets \FCM{X} and \FCM{Y} with labels of
pure best responses of the other player, and own labels
where a pure strategy has probability zero.}
\label{flhdiag}
\end{figure}

The two strategy sets $X$ and $Y$ with their subdivision
into best-response regions for the pure strategies of the
other player are now given additional labels at their
boundaries.
Namely, a point $x$ in $X$ gets label $i$ in
$\{1,\ldots,m\}$ if $x_i=0$, and a point $y$ in $X$ gets
label $m+j$ in $\{m+1,\ldots,m+n\}$ if $y_j=0$. 
That is, the ``outside labels'' correspond to a player's
\textit{own pure strategies that are played with probability
zero}.
Figure~\ref{flhdiag} shows this for the example
(\ref{example}).
A point may have several labels of a player, if it has
multiple best responses or more than one own strategy that
has probability zero.
For example, $x=(1,0,0)$ has the three labels $2,3,4$.
The points in $X$ that have three labels, and the points in
$Y$ that have two labels, are marked as dots in
Figure~\ref{flhdiag}. 
Apart from the unit vectors that are the vertices (corners)
of $X$ and $Y$, these are the points $a$ and $b$ in $X$ and
$c$ and $d$ in~$Y$.
With these labels, an equilibrium is any \textit{completely
labeled} pair $(x,y)$, that is, every label in
$\{1,\ldots,m+n\}$ is a label of $x$ or of $y$, as the
next proposition asserts.

\begin{proposition}
\label{p-completely}
Let $(x,y)\in X\times Y$ for an $m\times n$ bimatrix game
$(A,B)$.
Then $(x,y)$ is a Nash equilibrium of $(A,B)$ if and only if
$(x,y)$ is completely labeled.
\end{proposition}

\proof
A missing label would represent a pure strategy of either
player that is not a pure best response but has positive
probability, which is exactly what is not allowed in an
equilibrium according to Proposition~\ref{p-bestresponse}.
\endproof

The advantage of this condition is that it is purely
combinatorial and just depends on the labels but not on the
exact position of the dots in the diagrams in
Figure~\ref{flhdiag}.
There, because a completely labeled pair $(x,y)$ requires
all five labels, three of these must be labels of $x$ and
two must be labels of~$y$, so it suffices to consider the
finitely many points with these properties.
In $Y$, there are only four points $y$ that have two labels.
The first is $(1,0)$, which has labels 1 and~5.
There is indeed a point in $X$ which has the other labels 2,
3, 4, namely $(1,0,0)$, so $((1,0,0),(1,0))$ is an
equilibrium.
Point $d=(\frac23,\frac13)$ in $Y$ has labels 1 and 2, and
point $b$ in $X$ has the other labels 3, 4, 5, so
$(b,d)$ is another equilibrium, in agreement with
(\ref{bd}).
Point $c=(\frac13,\frac23)$ in $Y$ has labels 2 and 3, and
point $a$ in $X$ has the other labels 1, 4, 5, so
$(a,b)$ is a third equilibrium, in agreement with
(\ref{ac}).
Finally, point $(0,1)$ in $Y$ has labels 3 and 4, but there
is no point in $X$ that has the remaining labels 1, 2, 5, so
there is no equilibrium where player~2 plays $(0,1)$.
This suffices to identify all equilibria.
(The remaining points $(0,1,0)$ and $(0,0,1)$ of $X$ have
three labels, neither of which have corresponding points in
$Y$ that have the other two labels.) 

In the above example, no point in $X$ has more than three
labels, and no point in $Y$ has more than two labels.
In general, this is equivalent to the nondegeneracy of the
game.

\begin{proposition}
\label{p-nondeg-label}
An $m\times n$ bimatrix game is nondegenerate if and only if
no $x$ in $X$ has more than $m$ labels, and no $y$ in $Y$
has more than $n$ labels.
\end{proposition}

\proof
Let $x\in X$.
The labels of $x$ are the $|\br(x)|$ pure best responses
to~$x$ and player~1's own strategies $i$ where $x_i=0$,
where the number of the latter is $m-|\supp(x)|$.
So if the game is degenerate because $|\br(x)|>|\supp(x)|$, this is
equivalent to $|\br(x)|+m-|\supp(x)|>m$, that is,
$x$ having more than $m$ labels.
Similarly, $y$ in $Y$ has more that $|\supp(y)|$ pure best
responses if and only if $y$ has more than $n$ labels.
If this is never the case, the game is nondegenerate.
\endproof 

We need further concepts about polyhedra and polytopes. 
A \textit{polyhedron} $P$ in ${\mathbb R}^d$ is a set 
$\{z\in{\mathbb R}^d\mid Cz\le q\}$ for some matrix $C$ and vector~$q$.
It is called \textit{full-dimen\-sional} if it has dimension~$d$.
It is called a \textit{polytope} if it is bounded.
A~\textit{face} of $P$ is a set
$\{\,z\in P\mid c^{\top} z=q_0\}$ for some
$c\in {\mathbb R}^d$ and $q_0\in{\mathbb R}$
so that the inequality $c^{\top} z\le q_0$ is \textit{valid}
for $P$, that is, holds for all $z$ in~$P$.
A~\textit{vertex} of~$P$ is the unique element of a
0-dimensional face of~$P$.
An \textit{edge} of~$P$ is a one-dimensional face of~$P$.
A~\textit{facet} of a $d$-dimensional polyhedron $P$ is a
face of dimension~${d-1}$.
It can be shown that any nonempty face $F$ of $P$ can be obtained
by turning some of the inequalities that define~$P$ into
equalities, which are then called \textit{binding}
inequalities.
That is, $F=\{\,z\in P\mid c_i\T z=q_i\,,~i\in I\}$,
where $c_i\T z\le q_i$ for $i\in I$ are some of the rows in $Cz\le q$.
A~facet is characterized by a single binding inequality
$c_i\T z\le q_i$ which is \textit{irredundant}, that is
(after omitting any equivalent inequality), the
inequality cannot be omitted without changing the
polyhedron; the vector $c_i$ is called the \textit{normal
vector} of the facet.
A~$d$-dimensional polyhedron $P$ is called \textit{simple}
if no point belongs to more than $d$ facets of~$P$,
which is true if there are no special dependencies
between the facet-defining inequalities.

The subdivision of $X$ and $Y$ into best-response regions as
shown in the example in Figure~\ref{flhdiag} can be nicely
visualized for small games with up to four strategies per
player, because then $X$ and $Y$ have dimension at most
three. 
If the payoff matrix $A$ in the game $(A,B)$ has rows 
$a_1,\ldots,a_m$, then the best-response region for
player~1's strategy~$i$ is the set
$\{y\in Y\mid a_iy\ge a_ky,~k=1,\ldots,m\}$\,, which is a
polytope since $Y$ is bounded.
However, for general $m\times n$ games, the subdivision of
$Y$ and $X$ into best-response regions has more structure by
taking into account, as an additional dimension, the payoffs
$u$ and $v$ to player 1 and~2.
In Figure~\ref{f2br}, the upper envelope of expected payoffs
to player~1 is obtained by the smallest $u$ for the points
$(y,u)$ in $Y\times\reals$ so that
$3y_1+3y_2\le u$, 
$2y_1+5y_2\le u$, 
$0y_1+6y_2\le u$, 
or in general $Ay\le\1u$.
Similarly, Figure~\ref{f2br} shows the smallest $v$ for
$(x,v)$ in $X\times\reals$ with $3x_1+2x_2+3x_3\le v$ and
$2x_1+4x_2+0x_3\le v$, or in general $B\T x\le\1v$.
The \textit{best-response polyhedron} of a player is the set
of that player's mixed strategies together with the upper
envelope of expected payoffs (and any larger payoffs) to the
\textit{other} player.
The best-response polyhedra $\overline P$ and $\overline Q$ 
of players 1 and 2 are therefore 
\begin{equation}
\label{hedra}
\arraycolsep0pt
\begin{array}{rll}
   \overline P&{} =\{(x,v)\in{\mathbb R}^m\times {\mathbb R}
     &{}\mid x\ge\hbox{\bf0},~\hbox{\bf1}^{\top} x =1,~ B^{\top} 
    x\le \hbox{\bf1} v\,\},\\
   \overline Q&{}=\{(y,u)\in{\mathbb R}^n\times {\mathbb R}
   &{} \mid Ay\le \hbox{\bf1} u,~ y\ge\hbox{\bf0},~\hbox{\bf1}^{\top} y =1\,\}\,.\\
%
\end{array}
\end{equation} 
Both polyhedra are defined by $m+n$ inequalities (and one
additional equation).
Whenever one of these inequalities is \textit{binding}, we
give it the corresponding \textit{label} in
$\{1,\ldots,m+n\}$.
For example, if in the example (\ref{example}) the inequality
$3x_1+2x_2+3x_3\le v$ of $\overline P$ is binding, that is,
$3x_1+2x_2+3x_3=v$, this means that the first pure strategy
of player~2, which has label~4, is a best response.
The best-response region with label~4 is therefore the
\textit{facet} of $\overline P$ for this binding inequality,
projected to the mixed strategy set $X$ by ignoring the
payoff~$v$ to player~2, as seen in Figure~\ref{f2br}.
Facets of polyhedra are easier to deal with than
subdivisions of mixed-strategy simplices into best-response
regions.

The binding inequalities of any $(x,v)$ in $\overline P$
and $(y,u)$ in $\overline Q$ define labels as before, so
that equilibria $(x,y)$ are again identified as completely
labeled pairs in $X\times Y$.
The corresponding payoffs $v$ and $u$ are then on the
respective upper envelope (that is, smallest), for the
following reason:
For any $x$ in $X$, at least one component $x_i$ of $x$ is
nonzero, so in an equilibrium label~$i$ must appear as a
best response to $y$, which means that the $i$th inequality
in $Ay\le\1u$ is binding, that is, $(Ay)_i=u$, so $u$ is on
the upper envelope of expected payoffs in $\overline Q$ as
claimed;
the analogous statement holds for any nonzero component
$y_j$ of $y$ with label $m+j$.

The polyhedra $\overline P$ and $\overline Q$ in
(\ref{hedra}) can be simplified by eliminating the payoff
variables $u$ and~$v$, by defining the following polyhedra:
\begin{equation}
\label{defPQ}
\arraycolsep0pt
\begin{array}{rllrlrl}
  P&{}=\{\,x\in {\mathbb R}^m&{}\mid {}&x&{}\ge\hbox{\bf0},~& B^{\top} x&{}\le \hbox{\bf1}\},\\
  Q&{}=\{\,y\in {\mathbb R}^n&{}\mid {}&Ay&{}\le \hbox{\bf1},~& y&{}\ge\bf 0\}\,.
\end{array}
\end{equation}
We want $P$ and $Q$ to be polytopes, which is equivalent to
$v>0$ and $u>0$ for any $(x,v)\in\overline P$ and
$(y,u)\in\overline Q$, according to the following
proposition.

\begin{proposition}
\label{p-posbr}
Consider a bimatrix game $(A,B)$.
Then $P$ in $(\ref{defPQ})$ is a polytope if and only if the
best-response payoff to any $x$ in $X$ is always positive,
and $Q$ in $(\ref{defPQ})$ is a polytope if and only if the
best-response payoff to any $y$ in $Y$ is always positive.
\end{proposition}

\proof
We prove the statement for $Q$; the proof for $P$ is
analogous.
The best-response payoff to any mixed strategy $y$ is the
maximum entry of $Ay$, so this is not always positive if and
only if $Ay\le\0$ for some~$y\in Y$.
For such a $y$ we have $y\ge0$, $y\ne\0$, and $y\alpha\in Q$
for any $\alpha\ge0$, which shows that $Q$ is not bounded.
Conversely, suppose the best-response payoff $u$ to any $y$
is always positive.
Because $Y$ is compact and $\overline Q$ is closed, the
minimum $u'$ of $\{u\mid \exists y~:~(y,u)\in\overline Q\}$ exists, 
$u'>0$, and $u\ge u'$ for all $(y,u)$ in $\overline Q$.
Then the map
\begin{equation}
\label{project}
\overline Q\to Q-\{\0\},
\qquad
(y,u)\mapsto y\cdot \textstyle\frac 1 u
\end{equation} 
is a bijection with inverse
$z\mapsto (z\cdot \frac1{\1\T z}), \frac1{\1\T z})$
for $z\in Q-\{\0\}$.
Here, $\frac1{\1\T z}\ge u'$ and thus $\1\T z\le 1/u'$,
where $\1\T z$ is the 1-norm $\sum_{j=1}^n|z_j|$ of $z$
(because $z\ge\0$), which proves that $Q$ is bounded and
therefore a polytope.
\endproof

As a sufficient condition that $v>0$ and $u>0$ for any
$(x,v)$ in $\overline P$ and $(y,u)$ in~$\overline Q$, we
assume that
\begin{equation}
\label{ABpos} 
  \hbox{$A$ and $B^\top$ are nonnegative and have no zero
  column}
\end{equation}
(because then $B\T x$ and $Ay$ are nonnegative and nonzero
for any $x\in X$, ${y\in Y}$).
We could simply assume $A>\bf 0$ and $B>\bf 0$, but it is
useful to admit zero matrix entries (e.g.\ as in the identity
matrix).
Note that condition (\ref{ABpos}) is not necessary for
positive best-response payoffs (which is still the case, for
example, if the zero entry of $A$ in (\ref{example}) is
negative, as Figure~\ref{f3br} shows).
By adding a suitable positive constant to all payoffs of a
player, which preserves the preferences of that player, we
can assume (\ref{ABpos}) without loss of generality.

With positive best-response payoffs, the polytope $P$ is
obtained from $\overline P$ by dividing each inequality
$\sum_{i=1}^mb_{ij}x_i\le v$ by~$v$, which gives
$\sum_{i=1}^mb_{ij}(x_i/v)\le 1$, and then treating $x_i/v$
as a new variable that is again called~$x_i$ in~$P$.
Similarly, $\overline Q$ is replaced by $Q$ by dividing each
inequality in $Ay\le\1u$ by $u$.
In effect, we have normalized the expected payoffs on the
upper envelope to be~1,
and dropped the conditions $\1^{\top} x=1$ and $\1^{\top} y=1$
(so that $P$ and $Q$ have full dimension, unlike $\overline
P$ and $\overline Q$).
Conversely, nonzero vectors $x\in P$ and $y\in Q$ are
multiplied by $v=1/\1\T x$ and $u=1/\1\T y$ to turn them
into probability vectors.
The scaling factors $v$ and $u$ are the expected payoffs to
the other player.  

\begin{figure}[hbt]
\vskip2ex
\unitlength1mm
\strut\hskip4mm\strut
\raise4mm\hbox{%
\begin{picture}(0,0)(0,0)
\put(-2,62){$P$}
\put(16,23){\small$(0,0,0)$}
\put(-4,-4){$(\frac13,0,0)$}
\put(20,1){$(\frac14,\frac18,0)=b$}
\put(44,16){$(0,\frac14,0)$}
\put(44,45){$(0,\frac14,\frac16)=a$}
\put(16,63){$(0,0,\frac13)$}
\end{picture}
\includegraphics[width=45mm]{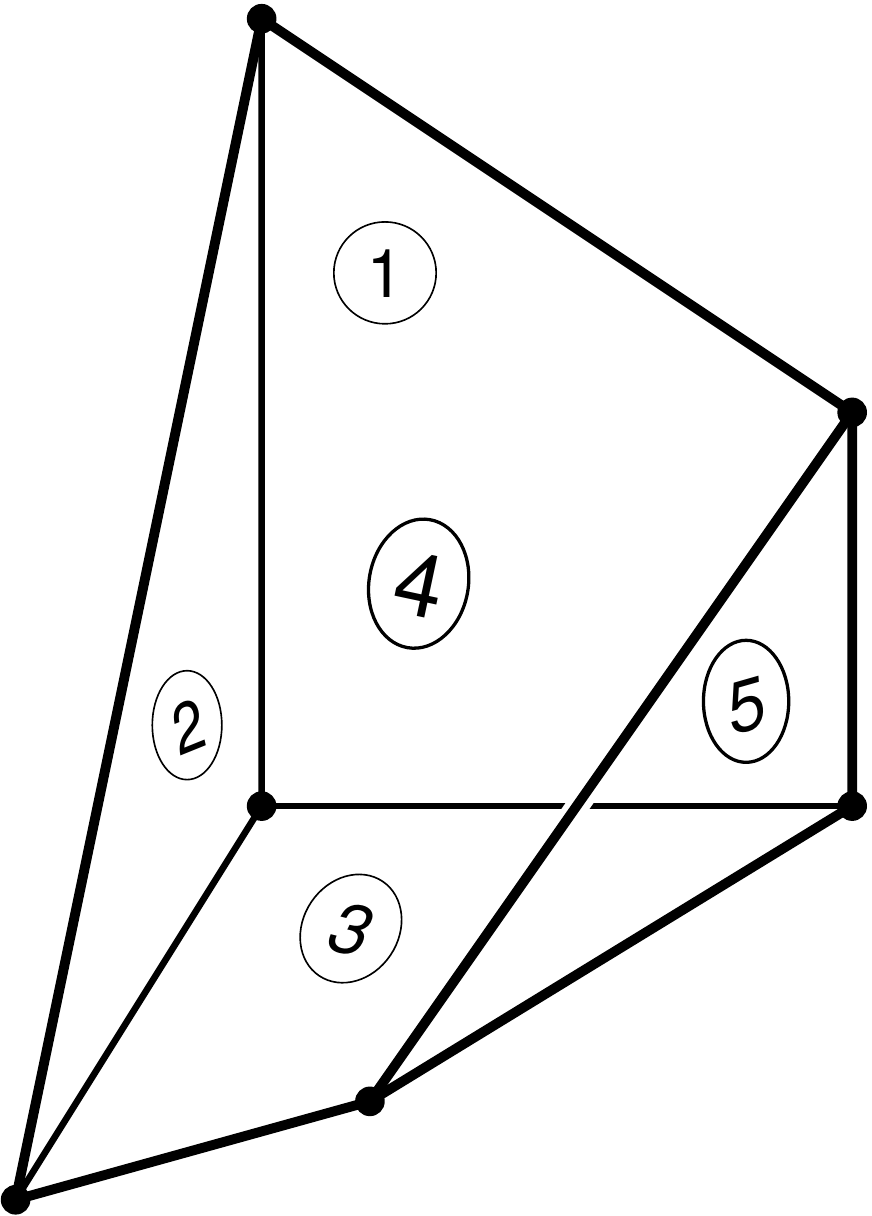}}
\hfill
\raise18mm\hbox{%
\begin{picture}(0,0)(0,0)
\put(0,48){$Q$}
\put(0,2){$(0,0)$}
\put(55,2){$(\frac13,0)$}
\put(44,26){$(\frac29,\frac19)=d$}
\put(20,36){$(\frac1{12},\frac16)=c$}
\put(0,36){$(0,\frac16)$}
\end{picture}
\includegraphics[width=60mm]{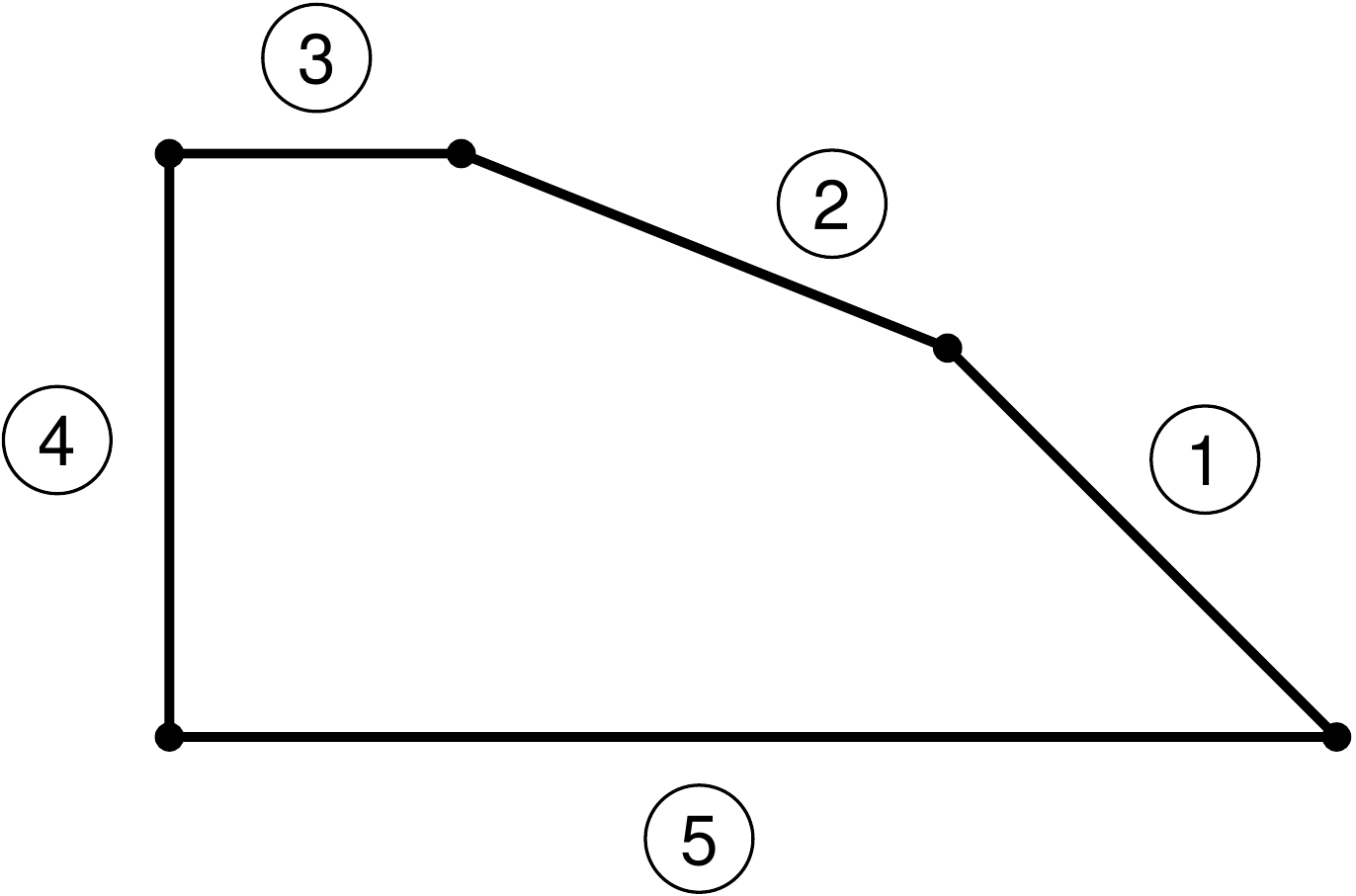}}
\strut\hskip3mm\strut
\caption{The polytopes \FCM{P} and \FCM{Q} in (\ref{defPQ})
for the game (\ref{example}).
Vertices are shown as dots, and facet labels as circled
numbers.}
\label{fPQ}
\end{figure}
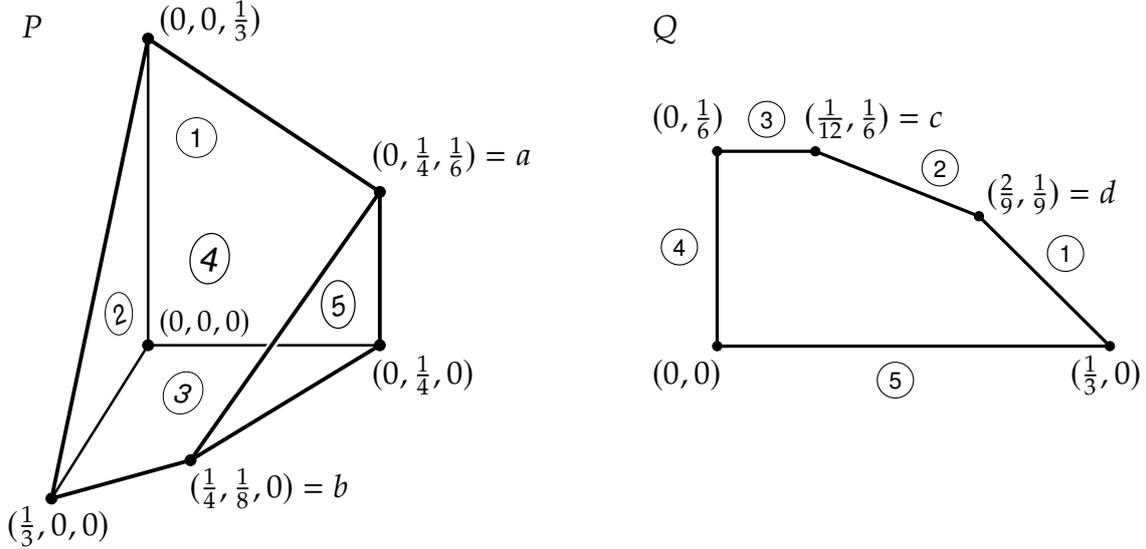 

Similar to (\ref{project}), the set $\overline P$ is in
one-to-one correspondence with $P-\{\0\}$ with the map
\begin{equation}
\label{project1}
\overline P\to P-\{\0\},
\qquad
(x,v)\mapsto x\cdot(1/v).
\end{equation} 
These bijections are not linear, but are known as ``projective
transformations''
(for a visualization see \citet*[Fig.~2.5]{vS2002}).
They map lines to lines, and any \textit{binding inequality}
in $\overline P$ (respectively, $\overline Q$)
corresponds to a binding inequality in $P$ (respectively,
$Q$) and vice versa.
Therefore, corresponding points have the same
\textit{labels} defined by the binding inequalities, which
are some of the $m+n$ inequalities that define $P$ and $Q$
in (\ref{defPQ}), see Figure~\ref{fPQ}.
An equilibrium is then a (re-scaled) completely labeled pair
$(x,y)\in P\times Q -\{(\0,\0)\}$ that has for each label
$i$ the respective binding $i$th inequality in $x\ge\0$ or
$Ay\le \1$, and for each label $m+j$ the respective binding
$j$th inequality in $B\T x\le\1$ or $y\ge\0$.

With assumption (\ref{ABpos}) and the polytopes $P$ and $Q$
in (\ref{defPQ}), an improved algorithm compared to
Algorithm~\ref{a-suppenum} is to find all completely
labeled \textit{vertex pairs} $(x,y)$ of $P\times Q$.
A simple example of the resulting improvement is an
$m\times 2$ game where $\overline Q$, similar to
Figure~\ref{f3br}, has at most $m+1$ vertices, as opposed to
about $m^2/2$ supports of size at most two for player~1.
For square games where $m=n$, the maximum number of
support pairs versus vertices to be tested changes from
about $4^n$ to about $2.6^n$
(\citealp{Keiding1997}; \citealp[p.~1724]{vS2002}).
We will describe an algorithm that finds all equilibria,
even in a degenerate game, in Section~\ref{s-maxnash}.
Before that, we will describe a classic algorithm that finds
at least one Nash equilibrium of a bimatrix game, which
also proves that a Nash equilibrium exists.

\section{The Lemke-Howson algorithm}
\label{s-lh}

\citet{LH} (LH) described an algorithm
that finds one Nash equilibrium of a bimatrix game.
It proves the existence of a Nash equilibrium for
nondegenerate games, which can also be adapted to degenerate
games.
We first explain this algorithm following
\cite*{Shapley1974}.
In the next section we describe it using the polytopes
from the previous Section~\ref{s-poly}.

Consider a nondegenerate bimatrix game and
Figure~\ref{flhdiag} for the example (\ref{example}).
The mixed-strategy simplices $X$ and $Y$ are subdivided into
best-response regions which are labeled with the other
player's best responses, and the facets of the simplices
are labeled with the unplayed own pure strategies.
These labels give rise to a \textit{graph} that consists of
finitely many vertices, joined by edges.
A vertex is any point of $X$ that has $m$ labels. 
An edge of the graph for $X$ is a set of points defined by
$m-1$ labels.
Its endpoints are the two vertices that have these ${m-1}$
labels in common; for example, the vertices $a$ and $b$ are
joined by the edge with labels 4 and~5.
As shown in Theorem~\ref{t-eq}(g) below, nondegeneracy
implies that the faces of $P$ that are defined by $m$ and
$m-1$ labels are indeed vertices and edges of $P$, which
correspond to these graph vertices and edges.
There are only finitely many sets with $m$ labels and
therefore only finitely many vertices.
Similarly, every vertex and edge of $Y$ is defined by $n$
and ${n-1}$ labels, respectively.

\begin{figure}[hbt]
\[
\begin{picture}(0,0)(0,0)
\put(55,96){$\tilde X$}
\put(310,68){$\tilde Y$}
\end{picture}
\includegraphics[width=150mm]{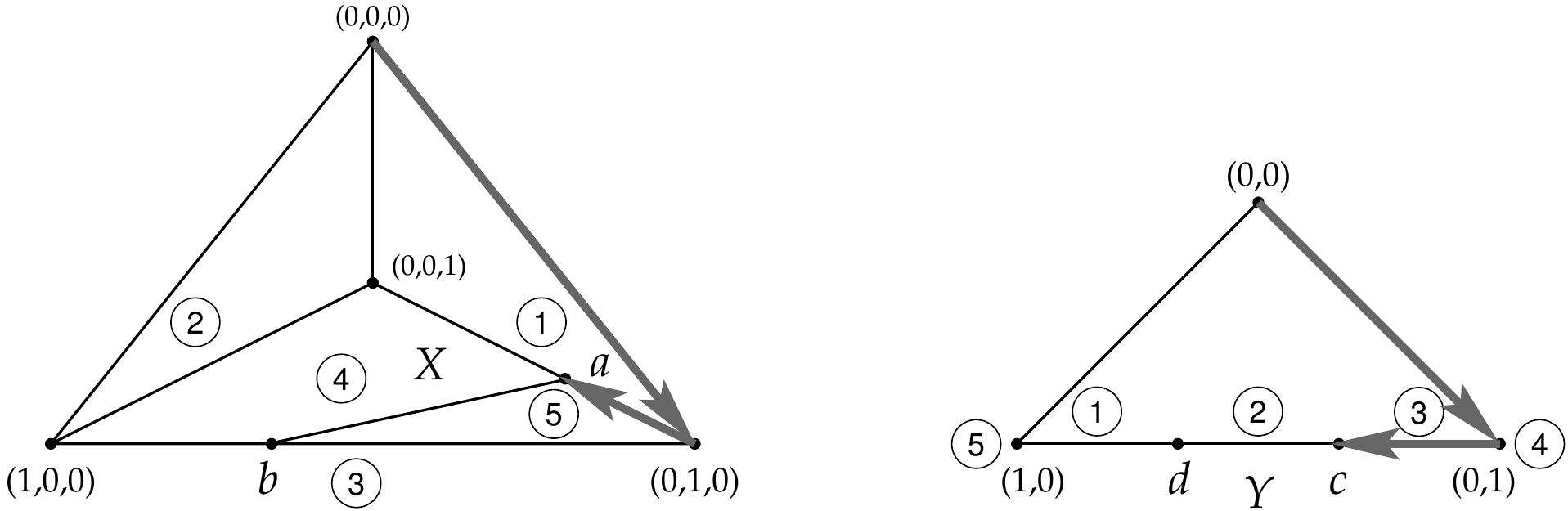}
\]
\caption{%
Extension \FCM{\tilde X} of \FCM{X} and \FCM{\tilde Y} of \FCM{Y}, each
with an additional vertex \FCM{\0}.
The artificial equilibrium \FCM{(\0,\0)} is completely labeled.
The arrows show the LH (Lemke-Howson) path starting from
\FCM{(\0,\0)} for missing label~2.
%
}
\label{flh}
\end{figure}
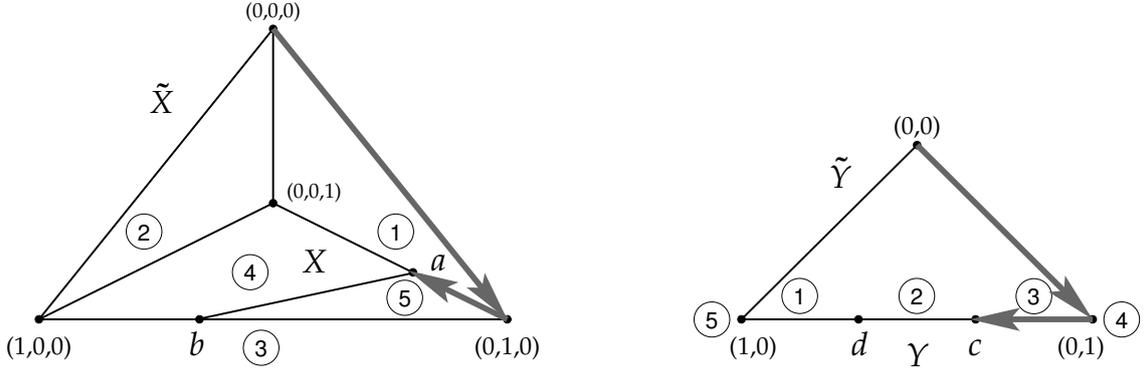

We now extend the graph for $X$ by adding another vertex
$\0$ in $\reals^m$ to obtain an extended graph $\tilde X$.
The new vertex $\0$ has all labels $1,\ldots,m$, and is
connected by an edge to each unit vector $e_i$ (which is a
vertex of $X$), which has all labels $1,\ldots,m$
except~$i$.
One can also consider $\tilde X$ geometrically as the convex
hull of $X$ and $\0$.
This is an $m$-dimensional simplex in $\reals^m$ with $X$ as
one facet (subdivided and labeled as before) and $m$
additional facets $\{x\in\tilde X\mid x_i=0\}$ for each
$i=1,\ldots,m$, with label~$i$, which produces the described
labels.
However, only the graph structure of $\tilde X$ matters.
In the same way, $Y$ is extended to~$\tilde Y$ with an extra
vertex $\0$ in $\reals^n$ that has all labels
$m+1,\ldots,m+n$, which is connected by $n$ edges to the $n$
unit vectors in $Y$.
The extended graphs $\tilde X$ and $\tilde Y$ are shown in
Figure~\ref{flh}.

The point $(\0,\0)$ of $\tilde X\times\tilde Y$ is
completely labeled, but does not represent a mixed strategy
pair.
We call it the \textit{artificial equilibrium}, which is the
starting point of the LH algorithm.
For the algorithm, one label $k$ in
$\{1,\ldots,m,m+1,\ldots,m+n\}$ is declared as possibly
\textit{missing}.
The algorithm computes a path in $\tilde X\times\tilde Y$,
which we first describe for our example and then in general.

Figure~\ref{flh} shows the LH path for missing label~2.
The starting point is $(x,y)=((0,0,0),(0,0))$, where $x$ has
labels $1,2,3$ and $y$ has labels $4,5$.
With label~2 allowed to be missing, we start by
\textit{dropping} label~2, which means changing $x$ along
the unique edge that connects $(0,0,0)$ to $(0,1,0)$
(shown by an arrow in the figure), while keeping $y=(0,0)$
fixed.
The endpoint $x=(0,1,0)$ of that arrow has a new label~5
which is \textit{picked up}.
Because $x$ has three labels $1,3,5$ and $y=(0,0)$ has two
labels $4,5$ but label~2 is missing, the label~5 that has just
been picked up in $\tilde X$ is now \textit{duplicate}.
Because $y$ no longer needs to have the duplicate label~5,
the next step is to drop label~5 in $\tilde Y$, that is,
change $y$ from $(0,0)$ to $(0,1)$ along the edge which has
only label~4.
At the end of that edge, $y=(0,1)$ has labels 4 and~3, where
label~3 has been picked up.
The current point $(x,y)=((0,1,0),(0,1))$ therefore has
duplicate label~3.
Correspondingly, we can now drop label~3 in $\tilde X$,
that is, move $x$ along the edge with labels 1 and~5 to
point~$a$, where label~4 is picked up.
At the current point $(x,y)=(a,(0,1))$, label~4 is
duplicate.
Next, we drop label~4 in $\tilde Y$ by moving $y$ along the
edge with label~3 to reach point~$c$, where label~2 is
picked up.
Because 2 is the missing label, the reached point
$(x,y)=(a,c)$ is completely labeled.
This is the equilibrium that is found as the endpoint of the
LH path.

In general, the algorithm traces a path that consists of
points $(x,y)$ in $\tilde X\times\tilde Y$ that have all
labels except possibly label~$k$.
Because $(x,y)$ has at least $m+n-1$ labels, this is only
possible in the following cases.
Suppose $x$ is a vertex of $\tilde X$ (which has $m$ labels)
and $y$ is a vertex of $\tilde Y$ (which has $n$ labels).
If $(x,y)$ has all labels $1,\ldots,m+n$ then it is an
equilibrium.
If $(x,y)$ has all labels except label~$k$, then $x$ and $y$
have exactly one label in common, which is the
duplicate label.
Alternatively, either $x$ has $m$ labels and is therefore a
vertex of $\tilde X$ and $y$ has $n-1$ labels and belongs to
an edge of $\tilde Y$, or $x$ has $m-1$ labels and belongs
to an edge of $\tilde X$ and $y$ has $n$ labels and is
therefore a vertex of $\tilde Y$.
These two possibilities $\{x\}\times F$ with $y\in F$
for an edge $F$ of $\tilde Y$,
or $E\times\{y\}$ with $x\in E$
for an edge $E$ of $\tilde X$, define the edges of the
\textit{product graph} $\tilde X\times \tilde Y$.
The vertices of this product graph are of the form $(x,y)$
where $x$ is a vertex of $\tilde X$ and $y$ is a vertex of
$\tilde Y$.
The LH algorithm generates a path in this product graph.
The steps of the algorithm alternate between traversing an
edge of $\tilde X$ while keeping a vertex of $\tilde Y$
fixed and vice versa.

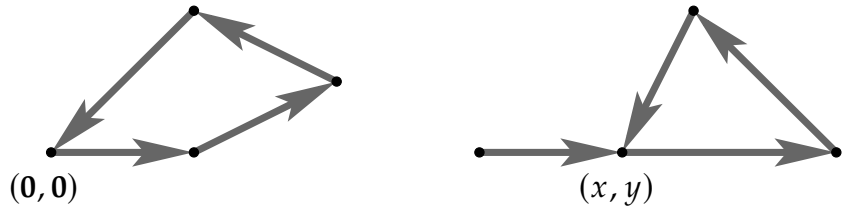
\begin{figure}[hbt]
\[
\unitlength1mm
\begin{picture}(0,0)(0,0)
\put(0,0){$(\0,\0)$}
\put(75,0){$(x,y)$}
\end{picture}
\includegraphics[width=110\unitlength]{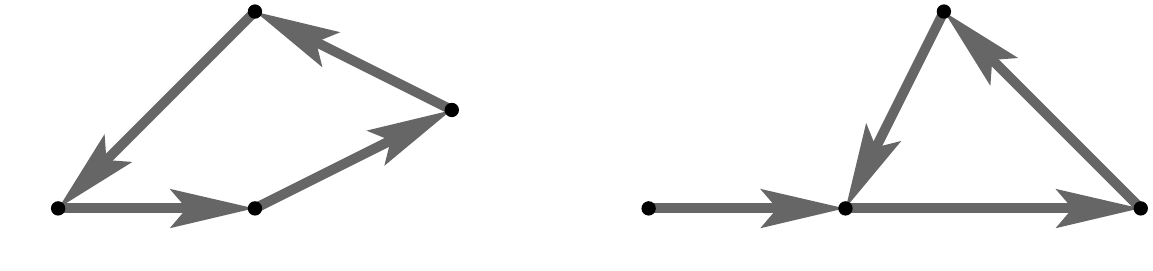}
\]
\caption{%
The LH algorithm cannot return to its starting point
\FCM{(\0,\0)} or re-visit an earlier vertex pair \FCM{(x,y)} because
this would imply alternative choices for starting or
continuing. 
}
\label{funique}
\end{figure}

The LH algorithm works because there is a unique next edge
in every step, which for the start depends on the chosen
missing label~$k$.
The algorithm starts from the artificial equilibrium
$(x,y)=(\0,\0)$ which is completely labeled.
If the missing label $k$ is in $\{1,\ldots,m\}$ then the
unique start is to move $x$ along the edge in $\tilde X$
that connects $\0$ to the unit vector $e_k$ because this is
the only edge that has all labels except~$k$.
If $k=m+j$ for $j$ in $\{1,\ldots,n\}$ then the unique start
is to move $y$ in $\tilde Y$ to $e_j$.
After that, a new label is picked which (unless it is~$k$)
is duplicate, and there is a unique edge in the other graph
($\tilde Y$ or $\tilde X$) where that duplicate label is
dropped, to continue the path.
If the label that is picked up is the missing label~$k$ then
the algorithm terminates at an equilibrium.
This cannot be the artificial equilibrium because the edge
that reaches the equilibrium would offer a second way to
start, which is not the case (because any edge of $\tilde
X\times\tilde Y$ that has all labels except $k$ could also be
traversed in the other direction).
Similarly, a vertex pair $(x,y)$ of $\tilde X\times\tilde Y$
cannot be re-visited because this would mean a second way to
continue, which is also not the case.
These two (excluded) possibilities are shown abstractly in
Figure~\ref{funique}.

\begin{figure}[hbt]
\[
\begin{picture}(0,0)(0,0)
\put(55,96){$\tilde X$}
\put(310,68){$\tilde Y$}
\end{picture}
\includegraphics[width=150mm]{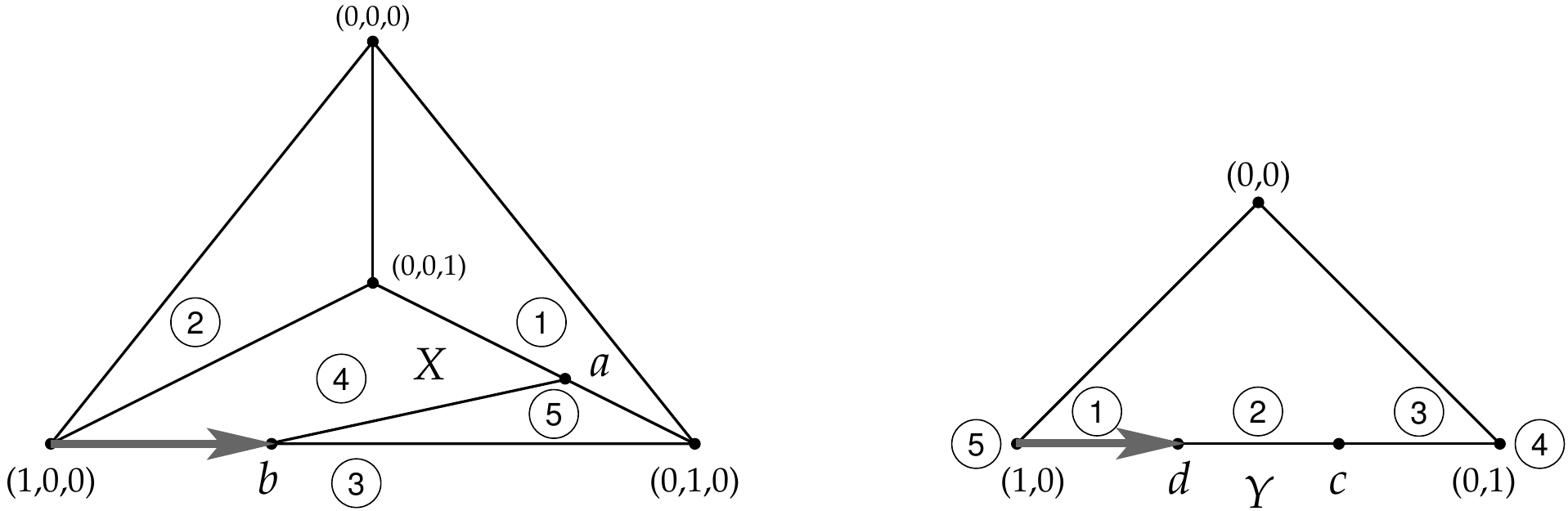}
\]
\caption{LH path for missing label 2 when started at the
pure-strategy equilibrium \FCM{((1,0,0),(1,0))}, which leads to
the equilibrium \FCM{(b,d)}.}
\label{flh-other}
\end{figure}
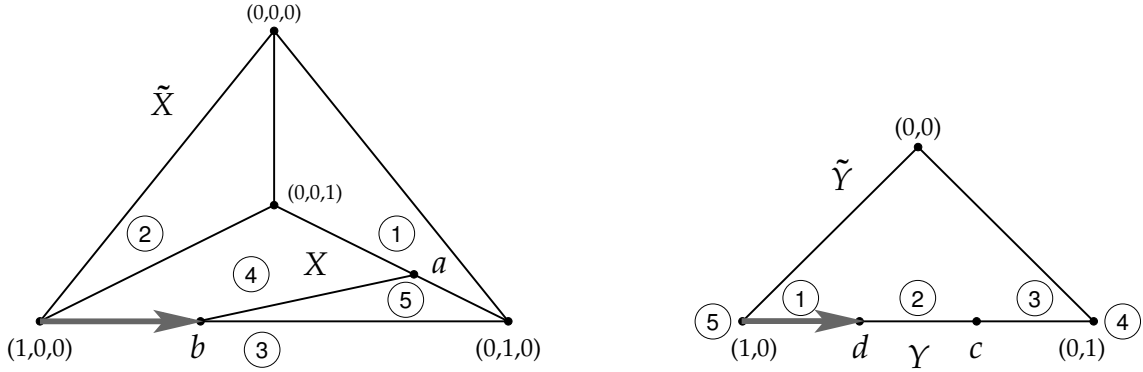

The LH algorithm can be started at any equilibrium, not just
the artificial equilibrium $(\0,\0)$.
For example, in Figure~\ref{flh}, starting it with missing
label~2 from the equilibrium $(a,c)$ that has just been
found would simply traverse the path back to the artificial
equilibrium.
However, as shown in Figure~\ref{flh-other}, if started from
the pure-strategy equilibrium $(x,y)=((1,0,0),(1,0))$ for missing
label~2, it proceeds as follows:
Dropping label~2 in $\tilde X$ changes to $(b,y)$ where
label~5 is picked up.
Dropping the duplicate label~5 in $\tilde Y$ changes to
$(b,d)$ where label~2 is picked up. 
This is the missing label, so the algorithm finds the
equilibrium $(b,d)$.
This has to be a new equilibrium because $(\0,\0)$ and
$(a,c)$ are connected by the unique LH path for missing
label~2 to which there is no other access.

Hence, we obtain the following important consequence.

\begin{theorem}
\label{t-odd}
Any nondegenerate bimatrix game has an odd number of Nash
equilibria.
\end{theorem}

\proof
Fix a missing label~$k$.
Then the artificial equilibrium $(\0,\0)$ and all Nash
equilibria are the unique endpoints of the LH paths for
missing label~$k$.
The number of endpoints of these paths is even, exactly one
of which is the artificial equilibrium, so the number of
Nash equilibria is odd.
\endproof

The LH paths for missing label~$k$ are the sets of edges and
vertices of $\tilde X\times\tilde Y$ that have all labels
except possibly~$k$.
These may also create cycles which have no endpoints.
Such cycles may occur but do not affect the algorithm.

A \textit{different} missing label may change how the 
artificial equilibrium and the Nash equilibria are ``paired''
as endpoints of each LH path for that missing label.
For example, any \textit{pure} Nash equilibrium is connected
in two steps to the artificial equilibrium via a suitable
missing label.
Suppose the pure strategy equilibrium is $(i,j)$.
Choose $k=i$ as the missing label.
Then the LH path first moves in $\tilde X$ to $(e_i,\0)$
where the label that is picked up is $m+j$ because $j$ is
the best response to~$i$.
The next step is then to $(e_i,e_j)$ where the algorithm
terminates because the best response to $j$ is $i$ which is
the missing label.
In the above example in Figure~\ref{flh-other}, the
pure-strategy equilibrium $((1,0,0),(1,0))$ can therefore be
found via missing label~1 (or missing label~4 which
corresponds to player~2's pure equilibrium strategy).
As shown earlier, missing label~2 connects the artificial
equilibrium to $(a,c)$, and therefore the LH path for
missing label~2 when started from $((1,0,0),(1,0))$
necessarily leads to a third equilibrium.
However, the ``network'' obtained by connecting equilibria
via LH paths for different missing labels may still not
connect all Nash equilibria directly or indirectly to the
artificial equilibrium.
An example due to Robert Wilson has been given by
\citet[Fig.~3]{Shapley1974}, which is a $3\times 3$ game
where for every missing label the LH path from the
artificial equilibrium leads to the same Nash equilibrium,
and two further Nash equilibria (which are unreachable this
way) are connected to each other.

In order to run the LH algorithm, it is not necessary to
create the graphs $\tilde X$ and $\tilde Y$ in full (which
would directly allow finding all Nash equilibria as
completely labeled vertex pairs).
Rather, the alternate traversal of the edges of these graphs
can be done in each step by a local ``pivoting'' operation
that is similarly known for the simplex algorithm for linear
programming.
We explain this in Section~\ref{s-pivot}.

\section{Lemke-Howson paths on polytopes}
\label{s-LHpoly}

A convenient way to implement the LH algorithm uses the
polytopes $P$ and $Q$ in (\ref{defPQ}) rather than the
projections of the best-response polyhedra 
$\overline P$ and $\overline Q$ in (\ref{hedra}) to $X$
and~$Y$.
The polytopes $P$ and $Q$ have the extra point $\0$ which is
the only point not in correspondence to the polyhedron 
$\overline P$ and $\overline Q$ via a projective
transformation as in (\ref{project}).
The extra point $(\0,\0)\in P\times Q$ is completely labeled
and represents the artificial equilibrium where the LH
algorithm starts.

We now consider a more general setting.
A \textit{Linear Complementarity Problem} or LCP
is given by a $d\times d$ matrix $C$ and a vector
$q\in\reals^d$, where the problem is to find $z\in\reals^d$
so that 
\begin{equation}
\label{LCP}
z\ge\0, \qquad w=q-Cz\ge\0, \qquad z\T w=0 
\end{equation}
(the standard notation for an LCP, see \citealp{CPS},
uses $-M$ instead of $C$ and $n$ instead of~$d$).
In (\ref{LCP}), because both $z$ and $w$ are nonnegative, the orthogonality
condition $z\T w=0$ is equivalent to the condition
$z_iw_i=0$ for each $i=1,\ldots,d$, which means that at least
one of the variables $z_i$ and $w_i$ is zero; these
variables are therefore also called \textit{complementary}.

A geometric way to view an LCP is the following.
Consider the polyhedron $S$ in $\reals^d$ given by
\begin{equation}
\label{S}
S=\{z\in\reals^d\mid z\ge\0,~Cz\le q\,\}.
\end{equation}
For any $z\in S$, we say $z$ has \textit{label}~$i$ in
$\{1,\ldots,d\}$ if $z_i=0$ or if $(Cz)_i=q_i$,
and call $z$ \textit{completely labeled} if
$z$ has all labels $1,\ldots,d$.
Clearly, $z$ is a solution to the LCP (\ref{LCP}) if and
only if $z\in S$ and $z$ is completely labeled.

In $S$, the $2d$ inequalities $z\ge0$, $Cz\le q$ have the
labels $1,\ldots,d,1,\ldots,d$ (which means every label
occurs twice) and $z$ in $S$ has label $i$ if one of the
corresponding inequalities is binding. 
The labels of $S$ in (\ref{S}) should be thought of as
labeling the \textit{facets} of~$S$.
We assume $S$ is \textit{nondegenerate}, that is, no $z\in
S$ has more than $d$ binding inequalities.
As shown in Theorem~\ref{t-eq}(h) below, this is
equivalent to the following
conditions: $S$ is a simple polytope (no point is on more
than $d$ facets), and no inequality can be omitted without
changing~$S$, unless it is never binding.
Every facet therefore corresponds to a unique binding
inequality, and has the corresponding label.
Any edge of $S$ is defined by $d-1$ facets, and any vertex
by $d$ facets.
Any point has the labels of the facets it lies on.

Consider an $m\times n$ bimatrix game $(A,B)$, which may
be degenerate.
Assume that $P$ and $Q$ in (\ref{defPQ}) are polytopes, if
necessary by adding a constant to the payoffs (see
Proposition~\ref{p-posbr}).
Then any Nash equilibrium of $(A,B)$ is given by a solution
$z=(x,y)\in P\times Q=S$ with $z\ne\0$ to the LCP
(\ref{LCP}).
That is, $d=m+n$ and $q=\1\in\reals^{m+n}$, and
\begin{equation}
\arraycolsep.0em
\label{C}
C= \left[
\begin{matrix}
0~~~ & A\\
B\T~~ & 0\\
\end{matrix}
\,\right]
\end{equation}
where $0$ is an all-zero matrix (of size $m\times m$ and
$n\times n$, respectively).
The $m+n$ labels are exactly as described in
Section~\ref{s-poly}, and correspond to unplayed pure
strategies~$i$ if $z_i=0$ or best-response pure
strategies~$i$ if $(Cz)_i=q_i=1$.
As before, for $z=(x,y)$ the vectors $x$ and $y$ have
to be re-scaled to represent mixed strategies. 
Moreover, $S$ is nondegenerate if and only if the game
$(A,B)$ is nondegenerate, by
Proposition~\ref{p-nondeg-label}.

We now study the LH algorithm without assuming the
product structure $S=P\times Q$ for $S$, which simplifies
the description.
Let $S$ in (\ref{S}) be a nondegenerate polytope 
so that $\0$ is a vertex of~$S$.
By nondegeneracy, when $z=\0$ then the remaining $d$
inequalities $Cz\le q$ are strict, that is, $\0<q$.
We can therefore divide the $i$th inequality (that is, the
$i$th row of $C$ and of~$q$) by $q_i$ and thus assume
$q=\1$.
This polytope has also a game-theoretic interpretation.

\begin{proposition}
\label{p-sym}
Let
\begin{equation}
\label{sym}
S=\{z\in\reals^d\mid z\ge\0,~Cz\le \1\,\}
\end{equation}
be a polytope with its $2d$ inequalities labeled
$1,\ldots,d,1,\ldots,d$.
Then $z\in S-\{\0\}$ is completely labeled if and only if
(with $z$ re-scaled as a mixed strategy)
$(z,z)$ is a symmetric Nash equilibrium of the symmetric
$d\times d$ game $(C,C\T)$. 
\end{proposition}

\proof
In the game $(C,C\T)$, let $y$ be a mixed strategy of
player~2, where the best-response payoff $\max_i (Cy)_i$
against $y$ is always positive because $S$ is a polytope 
(see Proposition~\ref{p-posbr} where this is stated for $Q$
instead of~$S$). 
Re-scaling the best-response payoff against $y$ to~$1$ and
re-scaling $y$ to $z$ gives the inequality $Cz\le\1$, where
$z\ge\0$.
By Proposition~\ref{p-bestresponse}, $z$ has all labels
$1,\ldots,d$ if and only if $(y,y)$ is a Nash equilibrium of
$(C,C\T)$. 
\endproof

Hence, the equilibria of a bimatrix game $(A,B)$ correspond
to the symmetric equilibria of the symmetric game $(C,C\T)$
in (\ref{C}).
This ``symmetrization'' seems to be a folklore result, first
stated for zero-sum games by \citet{GKT}.

We now express the LH algorithm in terms of computing a path
of edges of the polytope~$S$.
\def\ZZ{x}

\begin{proposition}
\label{p-S}
Suppose $S$ in $(\ref{sym})$ is a nondegenerate polytope,
with its $2d$ inequalities labeled $1,\ldots,d,1,\ldots,d$.
Then $S$ has an even number of completely labeled vertices,
including~$\0$.
\end{proposition}

\proof
This a consequence of the LH algorithm applied to $S$.
Fix a label $k$ in $\{1,\ldots,d\}$ as allowed to be missing
and consider the set of all points of $S$ that have all
labels except possibly~$k$.
This defines a set of vertices and edges of~$S$, which we
call the \textit{missing-$k$} vertices (which may
nevertheless also have label~$k$) and edges.
Any missing-$k$ vertex is either completely labeled (for
example, $\0$), or has a duplicate label, say~$\ell$.
A completely labeled vertex $z$ is the endpoint of a unique
missing-$k$ edge which is defined by the $d-1$ facets
that contain~$z$ except for the facet with label~$k$, by
``moving away'' from that facet.
If the missing-$k$ vertex $z$ does not have label~$k$, then
it is the endpoint of two missing-$k$ edges, each obtained
by moving away from one of the two facets with the duplicate
label~$\ell$.
Hence, the missing-$k$ vertices and edges define a
collection of paths and cycles, where the endpoints of the
paths are the completely labeled vertices.
Their total number is even because each path has two
endpoints.
\endproof

\begin{figure}[hbt]
\unitlength1mm
\begin{picture}(0,0)(0,0)
\put(5,3){$\0$}
\put(45,18){$\ZZ$}
\end{picture}
\includegraphics[height=60\unitlength]{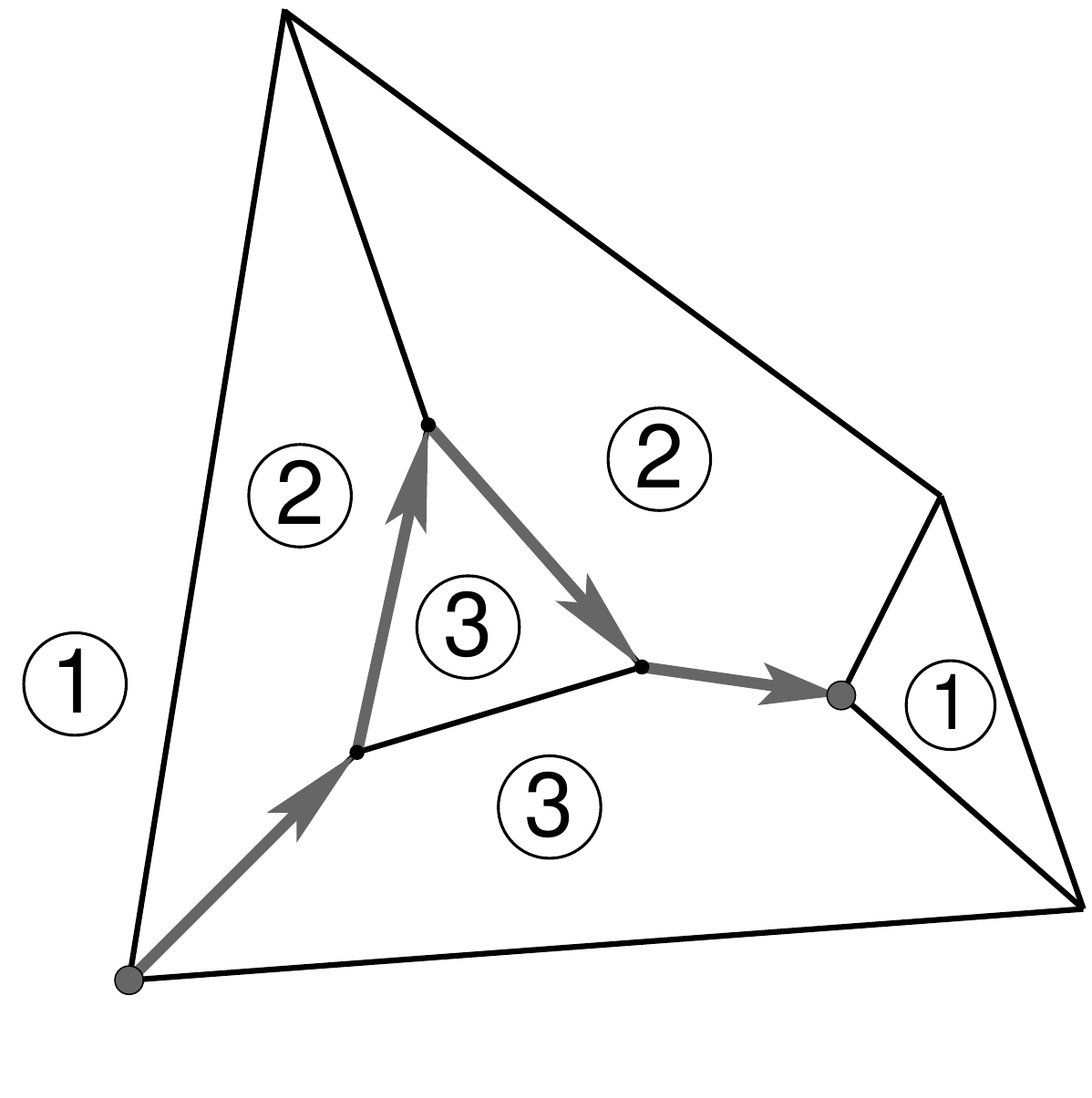}
\hfill
\begin{picture}(0,0)(0,0)
\put(12,1){$-1$}
\put(46,12){$+1$}
\put(5,3){$\0$}
\put(45,18){$\ZZ$}
\end{picture}
\includegraphics[height=60\unitlength]{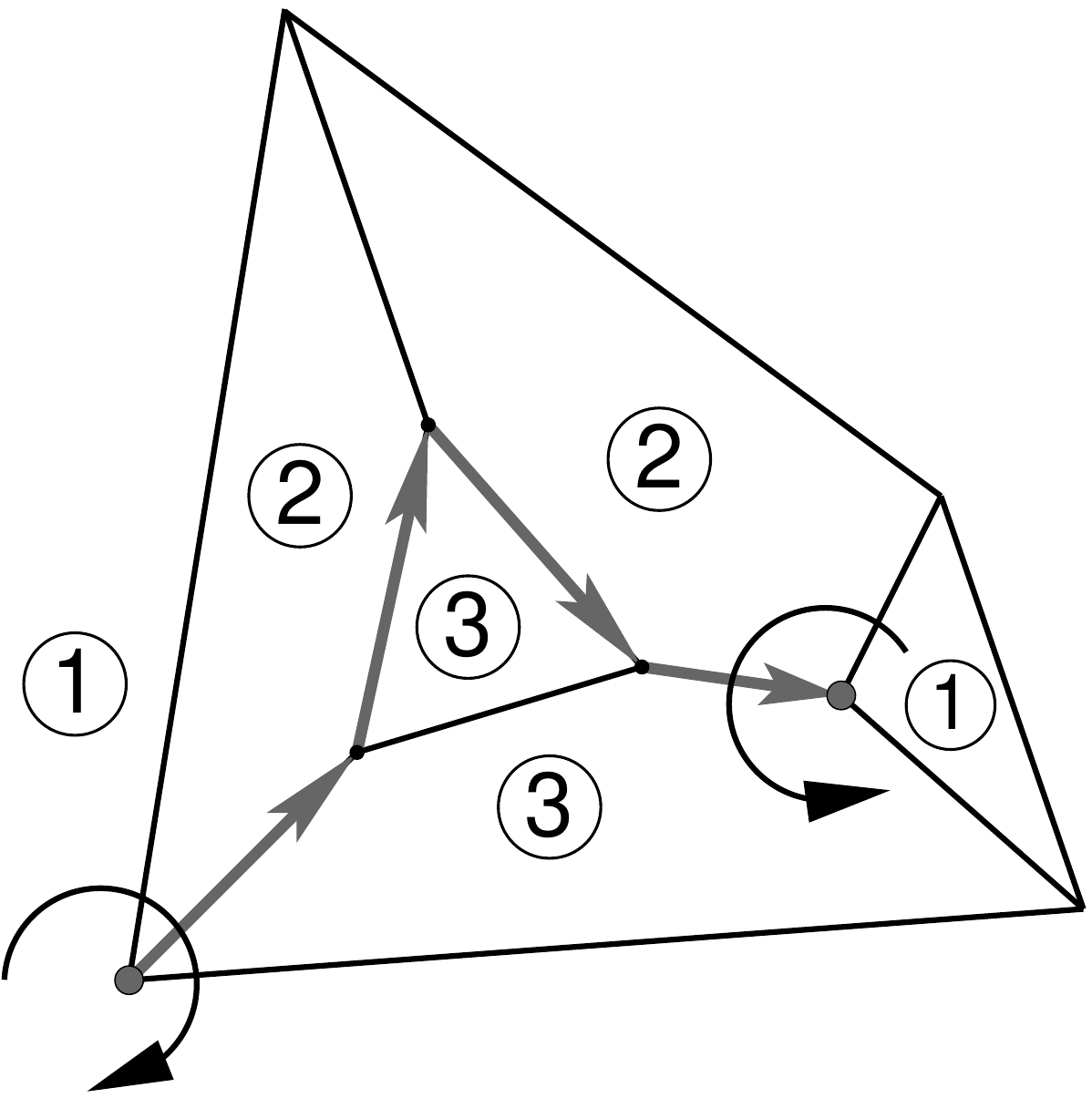}
\caption{Left: LH path for missing label 1 for the polytope
\FCM{S} with \FCM{C} as in (\ref{Spol}).
Right: Opposite orientation \FCM{-1} and \FCM{+1} of the labels
\FCM{1,2,3} around the two completely labeled vertices \FCM{\0} and
\FCM{\ZZ}.
}
\label{fpaths}
\end{figure}
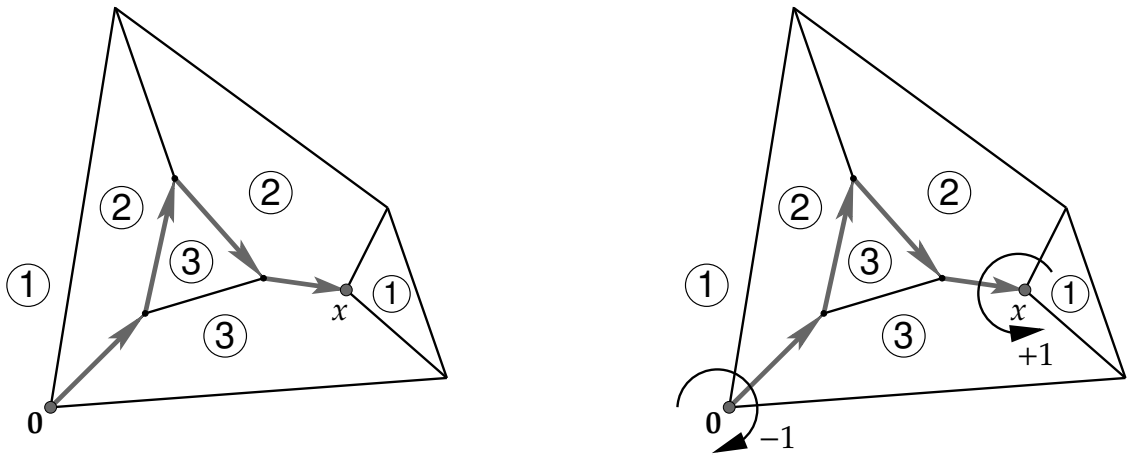

For the game $(C,C\T)$ with
\begin{equation}
\label{Spol}
\arraycolsep.4em
C 
= \left[\,\begin{matrix}
0&3&0\\
2&2&2\\
4&0&0
\end{matrix}\,\right],
\end{equation}
the polytope $S$ in (\ref{sym}) is shown in
Figure~\ref{fpaths} in a suitable planar projection (where
all facets are visible except for the facet defined by
$z_1=0$ with label~1 at the back of the polytope).
The diagram shows also the LH path for missing label~1.
The polytope has only two completely labeled vertices,
$\0$ and $\ZZ=(\frac16,\frac13,0)$.

\section{Endpoints of LH paths have opposite index}
\label{s-index}
In this section we prove a stronger version of
Proposition~\ref{p-S}.
Namely, the endpoints of an LH path will be shown to have 
opposite ``signs'' $-1$ and $+1$, which are independent of
the missing label.
This ``sign'' is called the \textit{index} of a Nash
equilibrium, which
we define here in an elementary way
using determinants.
By convention, the artificial equilibrium has index $-1$.
This implies that every nondegenerate bimatrix game has
$r$ Nash equilibria of index $+1$ and $r-1$ Nash equilibria
of index $-1$, for some integer $r\ge1$. 

The right diagram in Figure~\ref{fpaths} illustrates this
concept geometrically.
For a given completely labeled vertex, the index describes the 
orientation of the labels around the vertex.
Around $\0$ the labels $1,2,3$ appear clockwise, which is
considered a negative orientation and defines index $-1$,
whereas around the Nash equilibrium $\ZZ$ they appear
counterclockwise, which is a positive orientation and
defines index~$+1$.
We can also argue geometrically that the endpoints of an LH
path, here for missing label~1, have opposite index.
Starting from $\0$, the unique edge with missing label~1 has
label~2 on the left side of the path and label~3 on the
right side of the path.
As can be seen from the diagram, this holds for all
missing-1 edges when following the path.
The path terminates when it hits a facet with label~1,
which is now \textit{in front} of the edge of the path
which has label~2 on the left and label~3 on the right, so
the labels $1,2,3$ are in opposite orientation to the
starting point where label~1 is \textit{behind} the edge of
the path.
In addition, the LH path has a well-defined local
\textit{direction} that indicates where to go ``forward'' in
order to reach the endpoint with index $+1$, even if one
does not remember where one started: 
The forward direction has label~2 on the left and label~3 on
the right.

We show these properties of the index for labeled
polytopes $S$ as in Proposition~\ref{p-S} for general
dimension~$d$.
Our argument substantially simplifies the proof by
\citet{Shapley1974} who first defined the index for Nash
equilibria of bimatrix games. 

\begin{definition}
\label{d-index}
Consider a labeled nondegenerate polytope 
\begin{equation}
\label{Sine}
\{z\in{\mathbb R}^d\mid {c_j\T z\le q_j}\hbox{ for }j=1,\ldots,N\}
\end{equation}
where each inequality $c_j\T z\le q_j$ for $j=1,\ldots,N$ 
has some label in $\{1,\ldots,d\}$.
Consider a completely labeled vertex $\ZZ$ of $S$ where
$\lambda(i)$ indicates the inequality
$c_{\lambda(i)}\T\ZZ=q_{\lambda(i)}$ that is binding for $\ZZ$
and has label~$i$, for $i=1,\ldots,d$.
Then the \emph{index} of $\ZZ$ is defined as the sign of the
following determinant (multiplied by $-1$ if $d$ is even):
\begin{equation}
\label{ind}
(-1)^{d+1}\,\sign|c_{\lambda(1)}\cdots c_{\lambda(d)}|.
\end{equation}
\end{definition}

A $d\times d$ matrix formed by $d$ linearly independent
vectors in $\reals^d$ has a non\-zero determinant, but its
sign is only well defined for a specific order of these
vectors.
For a vertex of a nondegenerate polytope as in (\ref{Sine}),
the normal vectors $c_j$ of its binding inequalities are
linearly independent (see Theorem~\ref{t-eq}(e) below).
When the vertex is completely labeled, we write down these
normal vectors in the order of their labels, that is, for
$j=\lambda(i)$ for $i=1,\ldots,d$, and consider the
resulting $d\times d$ determinant in~(\ref{ind}).
The sign correction for even dimension $d$ is made for the
following reason.
For the polytope $S$ in (\ref{sym}) we write $z\ge\0$ as
$-z\le\0$ so all inequalities go in the same direction as
required in (\ref{Sine}).
For the completely labeled vertex $\0$ we thus obtain
the determinant of the negative of the $d\times d$ identity
matrix, which is $1$ if $d$ is even and $-1$ if $d$ is odd.
In order to obtain a negative index for this artificial
equilibrium, we therefore multiply the sign of the
determinant with $(-1)^{d+1}$.

\begin{figure}[hbt]
\[
\unitlength1mm
\begin{picture}(0,0)(0,0)
\put(7,12){$x$}
\put(28,12){$y$}
\put(-5,14){$c_{\lambda(1)}$}
\put(14,22){$c_{\lambda(2)}$}
\put(14,3){$c_{\lambda(3)}$}
\put(37,10){$c_{\lambda(0)}$}
\end{picture}
\includegraphics[height=26\unitlength]{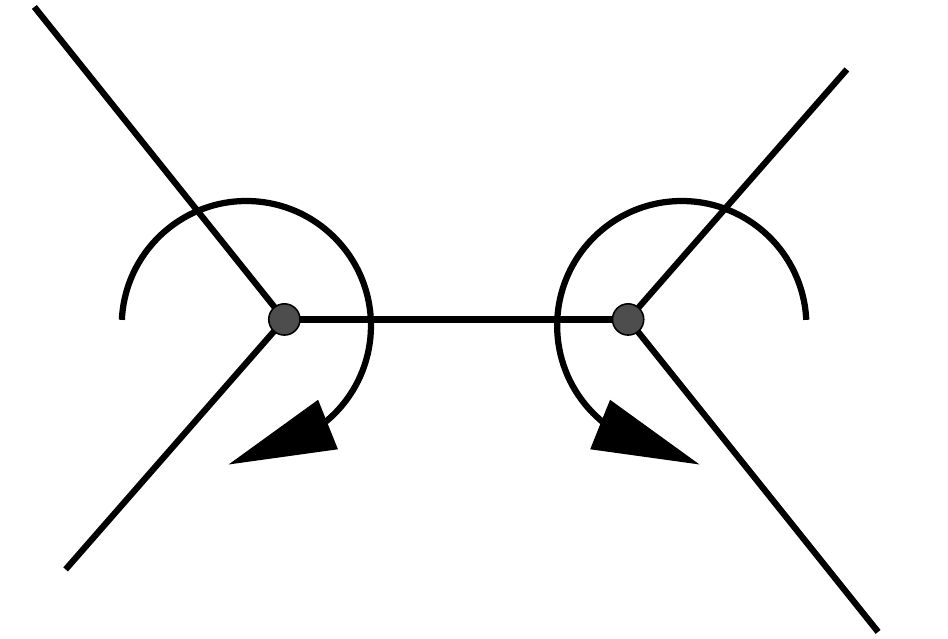}
\]
\caption{Opposite geometric orientation of adjacent vertices
\FCM{x} and~\FCM{y} as in Lemma~\ref{l-opp} for \FCM{d=3}.
The four involved facets are shown with their normal
vectors.
} 
\label{fopp}
\end{figure}
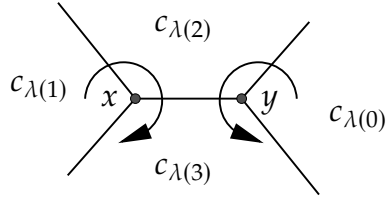

The next lemma states that ``pivoting changes sign'' in the
following sense.
``Pivoting'' is the algebraic representation of moving from
a vertex to an ``adjacent'' vertex along an edge.
This means that one binding inequality is replaced by
another.
For any fixed order of the normal vectors of the binding
inequalities, one of these vectors is thus replaced by
another, which we choose to be the vector in first position.
The lemma states that the corresponding determinants then
have opposite sign; it is geometrically illustrated in
Figure~\ref{fopp}.

\begin{lemma}
\label{l-opp}
Consider a nondegenerate polytope $S$ as in $(\ref{Sine})$,
and an edge defined by $d-1$ binding inequalities
$c_{\lambda(i)}\T \,z=q_{\lambda(i)}$ for $i=2,\ldots,d$.
Let $x$ and $y$ be the endpoints of this edge,
with the additional binding inequality 
$c_{\lambda(1)}\T \,x=q_{\lambda(1)}$ for~$x$ and
$c_{\lambda(0)}\T \,y=q_{\lambda(0)}$ for~$y$.
Then 
\begin{equation}
\label{detxy}
\sign|c_{\lambda(1)}c_{\lambda(2)}\ldots c_{\lambda(d)}|
=-
\sign|c_{\lambda(0)}c_{\lambda(2)}\ldots c_{\lambda(d)}|
\ne 0.
\end{equation}
\end{lemma}

\proof
Because $S$ is nondegenerate, $x$ and $y$ have exactly $d$
binding inequalities, so that the following conditions hold: 
\arraycolsep.2em
\begin{equation}
\label{stack}
\begin{array}{rclrcl}
c_{\lambda(0)}\T\, x&<&q_{\lambda(0)}~,
&\qquad
c_{\lambda(0)}\T\, y&=&q_{\lambda(0)}~,
\\
c_{\lambda(1)}\T\, x&=&q_{\lambda(1)}~,
&\qquad
c_{\lambda(1)}\T\, y&<&q_{\lambda(1)}~,
\\
c_{\lambda(2)}\T\, x&=&q_{\lambda(2)}~,
&\qquad
c_{\lambda(2)}\T\, y&=&q_{\lambda(2)}~,
\\
&\vdots&&&\vdots&\\
c_{\lambda(d)}\T\, x&=&q_{\lambda(d)}~,
&\qquad
c_{\lambda(d)}\T\, y&=&q_{\lambda(d)}~.
\\
\end{array}
\end{equation}
The $d+1$ vectors 
$c_{\lambda(0)},c_{\lambda(1)},\ldots,c_{\lambda(d)}$
are linearly dependent, so there are reals 
$\gamma_0,\gamma_1,\ldots,\gamma_d$, not all zero, with
\begin{equation}
\label{dep}
\gamma_0\,c_{\lambda(0)}\T
+
\gamma_1\,c_{\lambda(1)}\T
+\cdots+
\gamma_d\,c_{\lambda(d)}\T
=\0\T
\end{equation}
where $\gamma_0\ne0$ and $\gamma_1\ne0$ because otherwise
the binding inequalities for $x$ or $y$ would be linearly
dependent, which is not the case.
Hence, by (\ref{dep}) and (\ref{stack}),
\begin{equation*}
0=\0\T(y-x)=
\gamma_0\,c_{\lambda(0)}\T(y-x) + \gamma_1\,c_{\lambda(1)}\T(y-x)
\end{equation*}
and therefore
\begin{equation}
\label{ineq}
\frac{\gamma_0}{\gamma_1}~=~
\frac
{c_{\lambda(1)}\T \,x-c_{\lambda(1)}\T \,y}
{c_{\lambda(0)}\T \,y-c_{\lambda(0)}\T \,x}
~>~0
\end{equation}
by the first two rows in (\ref{stack}).
The linear dependence (\ref{dep}) and multilinearity of the
determinant imply
\begin{equation}
\label{det1}
\begin{array}{rcl}
0&=&|(
c_{\lambda(0)}\gamma_0+
c_{\lambda(1)}\gamma_1)
~
c_{\lambda(2)}
\cdots
c_{\lambda(d)}
| 
\\
&=&
| c_{\lambda(0)}
~
c_{\lambda(2)}
\cdots
c_{\lambda(d)}
| 
\,\gamma_0~+~
| c_{\lambda(1)}
~
c_{\lambda(2)}
\cdots
c_{\lambda(d)}
| 
\,\gamma_1 
\\
\end{array} 
\end{equation}
and thus
\begin{equation}
\label{det2}
| c_{\lambda(1)} ~ c_{\lambda(2)} \cdots c_{\lambda(d)} | 
~=~-
| c_{\lambda(0)} ~ c_{\lambda(2)} \cdots c_{\lambda(d)} | 
\frac{\gamma_0}{\gamma_1}
\end{equation}
which shows (\ref{detxy}). 
\endproof

\begin{theorem}
\label{t-opp}
Suppose $S$ in $(\ref{sym})$ is a nondegenerate polytope,
with its $2d$ inequalities labeled $1,\ldots,d,1,\ldots,d$.
Then $S$ has an even number of completely labeled vertices.
Half of these (including~$\0$) have index $-1$, the other 
half index $+1$.
The endpoints of any LH path have opposite index.
\end{theorem}

\proof
Let $S$ be described as in (\ref{Sine}), so that
$c_i=-e_i$ for $i=1,\ldots,d$ and
$C\T=[c_{d+1}\cdots c_{2d}]$.
Consider some completely labeled vertex $\ZZ$ of~$S$.
Let the binding inequalities for $\ZZ$ be
$c_{\lambda(i)}\T\,\ZZ=q_i$ with label~$i$ for $i=1,\ldots,d$.
We consider the LH path with missing label~1 that starts at
$\ZZ$, and show that the endpoint of that path has opposite
index to~$\ZZ$.
Suppose that $\ZZ$ has negative index and that $d$ is odd,
so that
$| c_{\lambda(1)} ~ c_{\lambda(2)} \cdots c_{\lambda(d)} |<0$.
On the first edge of the LH path with missing label~1, the
same inequalities as for $\ZZ$ are binding,  
except for the inequality $c_{\lambda(1)}\T\,z\le q_1$.
Let the endpoint of that edge be $y$, where now the
inequality $c_{\lambda(0)}\T\,y\le q_0$ is binding.
This is the situation of Lemma~\ref{l-opp}, so
$| c_{\lambda(0)} ~ c_{\lambda(2)} \cdots c_{\lambda(d)} |>0$
by (\ref{detxy}).
If the binding inequality $c_{\lambda(1)}\T\,y\le q_1$ has
the missing label~1, the claim is proved, because then $y$
is the other endpoint of the LH path, and has positive
index.

So suppose this is not the case, that is, the binding
inequality $c_{\lambda(1)}\T\,y\le q_1$ has a duplicate
label $\ell$ in $\{2,\ldots,d\}$.
We now \textit{exchange} columns $1$ and~$\ell$ in the
matrix 
$[ c_{\lambda(0)} ~ c_{\lambda(2)} \cdots c_{\lambda(d)} ]$,
which changes the sign of its determinant, which is now
\begin{equation}
\label{exch}
| c_{\lambda(\ell)} ~ c_{\lambda(2)} \cdots
c_{\lambda(\ell-1)}
~ c_{\lambda(0)} 
~ c_{\lambda(\ell+1)}
\cdots
c_{\lambda(d)} |
\end{equation}
and again negative.
Note that these are still the same normal vectors of the
binding inequalities for $y$, except for the exchanged
columns $1$ and~$\ell$; moreover, columns $2,\ldots,d$ have
labels $2,\ldots,d$ in that order.
The first column in (\ref{exch}) has the duplicate
label~$\ell$, and corresponds to the inequality
$c_{\lambda(\ell)}\T\,y\le q_\ell$ that is no longer binding 
when label $\ell$ is dropped for the next edge on the LH
path.
That is, (\ref{exch}) represents the same situation as the
starting point $\ZZ$\,:
The determinant is negative, columns $2,\ldots,d$ have the
correct labels, and the first column will be exchanged for a
new column when traversing the next edge.
The resulting determinant with the new first column has
opposite sign by Lemma~\ref{l-opp}.
If the label that has been picked up is the missing label~1,
then it is the endpoint of the LH path and the claim is
proved.
Otherwise we again exchange the first column with the column
of the duplicate label, with the determinant going back to
negative, and repeat, until the endpoint of the path is
reached.

On any missing-1 vertex on the path, we also can identify
the direction of the path by considering the two
determinants obtained by exchanging the first column and 
the column with its duplicate label (in both cases, columns
$2,\ldots,d$ have the correct labels).
The pivoting step (which determines the edge to be
traversed) replaces the first column of the determinant.
If it starts from a negative determinant, then the
direction is towards the endpoint with positive index (for
odd $d$, as in our description so far).
If it starts from a positive determinant, then the
direction is towards the endpoint with negative index.

Clearly, the analogous reasoning applies if the considered
starting point $\ZZ$ of the LH path has positive index or if
$d$ is even.
Because the endpoints of the LH paths for missing label~1
have opposite index, half of these endpoints have index $-1$
and the other half index $+1$, as claimed.

As concerns missing labels $k$ other than label~1, we can
reduce this to the case $k=1$ as follows:
We exchange the first and $k$th coordinate of $\reals^d$ as
the ambient space of $S$, and the first and $k$th row in the
$d$ inequalities $-z\le\0$ as well as in $Cz\le\1$.
This double exchange of rows and columns does not change the
signs of the determinant (\ref{ind}) of any completely
labeled vertex, and the LH path for missing label~$k$
becomes the LH path for missing label~1 where the preceding
reasoning applies. 
\endproof

\begin{figure}[hbt]
\unitlength1mm
\begin{picture}(0,0)(0,0)
\put(5,3){$\0$}
\put(21,16){$y$}
\put(3,23){$-e_1$}
\put(14,31){$-e_2$}
\put(29,14){$-e_3$}
\put(25,25){$c_3$}
\put(36,33){$c_2$}
\put(51,22){$c_1$}
\end{picture}
\includegraphics[height=60\unitlength]{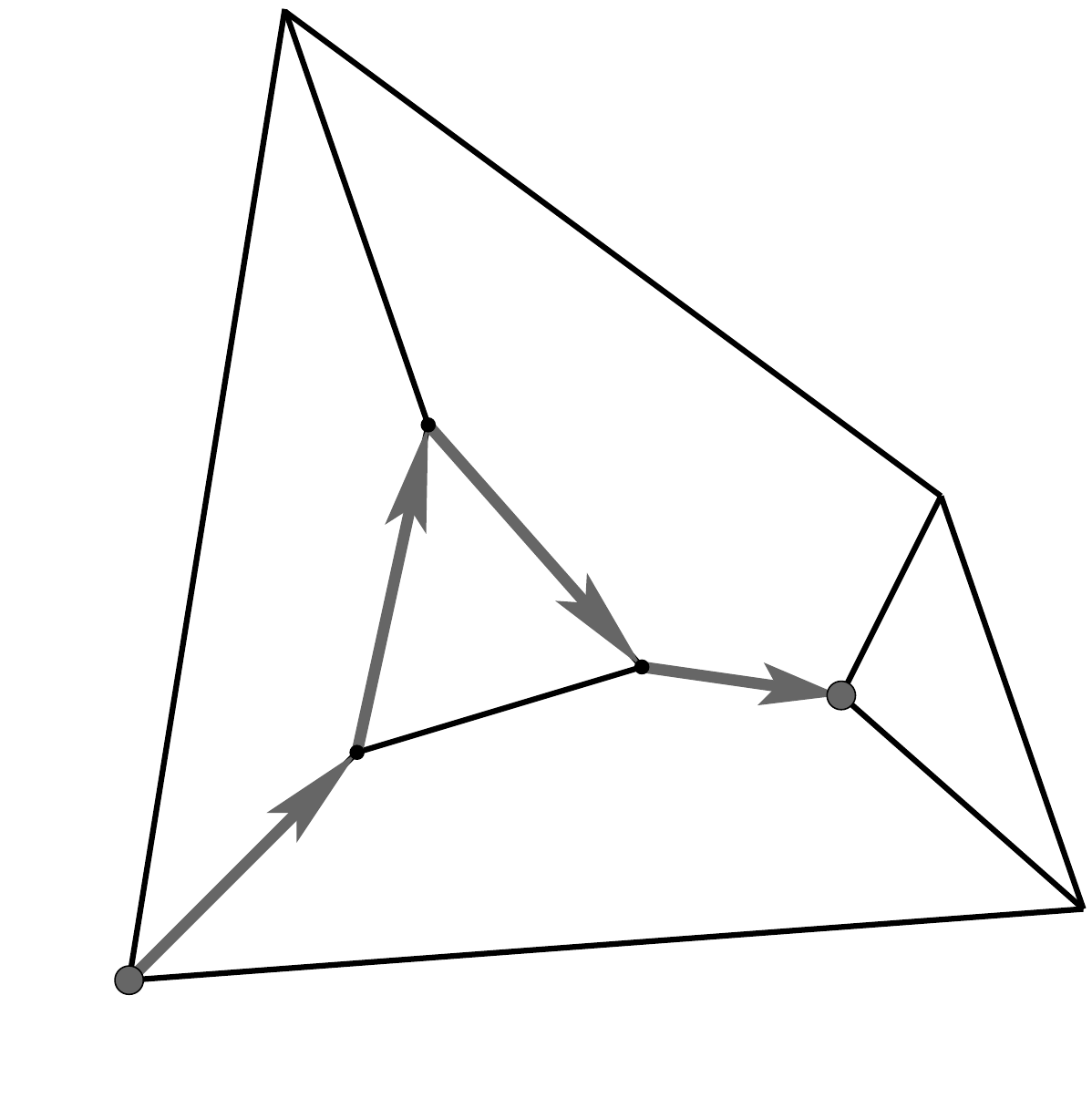}
\hfill
\raise 19ex\hbox{%
\begin{minipage}{.5\hsize}
\[
\arraycolsep.15em
\begin{array}{rrrrl c rrrrl}
 &      & -1  & & &
 &  &   & +1   & &\\
|& -e_1 & -e_2 & -e_3 & |
&~~\includegraphics[width=12mm]{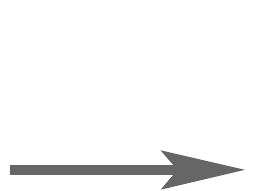}~~&
| & c_3 & -e_2 & -e_3 & |
\\
|& -e_3 & -e_2 & c_3 & |
& \includegraphics[width=12mm]{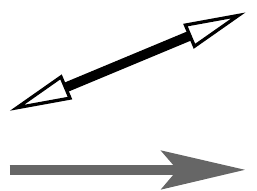} &
| & c_2 & -e_2 & c_3 & |
\\
|& -e_2 & c_2 & c_3 & |
& \includegraphics[width=12mm]{FIG/2arrow.pdf} &
|& -e_3 & c_2 & c_3 & |
\\
|& c_3 & c_2 & -e_3 & |
& \includegraphics[width=12mm]{FIG/2arrow.pdf} &
|& c_1 & c_2 & -e_3 & |
\\
\end{array}
\]
\end{minipage}}
\caption{Illustration of the proof of Theorem~\ref{t-opp}
with facets of \FCM{S} in (\ref{sym}) shown by their normal
vectors (whose subscripts are their labels), which are
either the negative unit vectors \FCM{-e_1}, \FCM{-e_2}, \FCM{-e_3}, or
\FCM{c_1\T}, \FCM{c_2\T}, \FCM{c_3\T} as the three rows of the matrix
\FCM{C} in~(\ref{Spol}).
%
}
\label{fdeter}
\end{figure}

Figure~\ref{fdeter} illustrates the proof of
Theorem~\ref{t-opp} where the right-hand side shows two
columns that display the determinants with sign $-1$ and
$+1$ for the steps of the algorithm.
It starts at the completely labeled vertex $\0$, which has
the negative unit vectors as normal vectors of its binding
inequalities.
Dropping label~1 and picking up label~3 exchanges $-e_1$
with~$c_3$, with the sign of the determinant changed from
$-1$ to $+1$.
This is the edge from $\0$ to~$y$.
The double-headed arrow shows the switch to the next line
which exchanges the columns $c_3$ and $-e_3$ with the
duplicate label~3, and brings the determinant back to $-1$,
but still refers to the same point~$y$.
The next step away from $y$ exchanges the column
$-e_3$ with~$c_2$ (it is always the first column that is
being replaced), and so on.
The last column that is found is $c_1$ which has the
missing label~1, with a positive determinant
$|c_1~c_2~{-}e_3|$ and hence positive index of the found Nash
equilibrium.

\section{Nondegenerate bimatrix games}
\label{s-nondegen}

Nondegeneracy of a bimatrix game $(A,B)$ is an important
assumption for the algorithms that we have described so far.
In Algorithm~\ref{a-suppenum}, which finds all equilibria by
support enumeration, it ensures that the equations that
define the mixed strategy probabilities for a given support
pair have unique solutions.
For the LH algorithm, it is, in addition, important for the
vertex pairs encountered on the LH path so that the path is
well defined.

The following theorem states a number of equivalent
conditions of nondegeneracy.
Some of them have been stated only as sufficient
conditions (but they are not stronger),
for example condition (e) by \citet[p.~52]{vandamme-book2}
and \citet{LH},
or (g) by \citet{Krohn} and, in slightly weaker form,
by \citet{Shapley1974}.
The purpose of this section is to state and prove the
equivalence of these conditions, which has not been done in
this completeness before.
Much of the proof is straightforward linear algebra, but
illustrative in this context, for example for the
implication (d) $\Rightarrow$~(e).
We comment on the different conditions afterwards.

\begin{theorem}
\label{t-eq}
Let $(A,B)$ be an $m\times n$ bimatrix game so that $P$ and
$Q$ in $(\ref{defPQ})$ are polytopes.
Consider $C$ as in $(\ref{C})$ where $d=m+n$ and
$C\T=[c_1\cdots c_d]$, and the polytope $S$ in
$(\ref{sym})$.
As before, a point in $P$, $Q$, or $S$ has label $i$ in
$\{1,\ldots,d\}$ if the corresponding $i$th inequality is
binding (in $S$ this can occur twice, for $z_i=0$ or $c_i\T
z=1$).
Then the following are equivalent.
\abs{\rm(a)}
$(A,B)$ is nondegenerate.
\abs{\rm(b)}
No point in $P$ has more than $m$ labels, and no point in
$Q$ has more than $n$ labels.
\abs{\rm(c)}
The symmetric game $(C,C\T)$ is nondegenerate.
\abs{\rm(d)}
For no point $z$ in $S$ more than $d$ of the inequalities
$z\ge\0$ and $Cz\le\1$ are binding.
\abs{\rm(e)}
For every $z\in S$ the row vectors $e_i\T$ and $c_j\T$ for
the binding inequalities of $z\ge\0$ and $Cz\le\1$ are
linearly independent. 
\abs{\rm(f)}
Consider any $\hat x\in X$ and $\hat y\in Y$.
Let $I=\supp(\hat x)$ and $J=\br(\hat x)$, and let
$B_{IJ}$ be the $|I|\times|J|$ submatrix of $B$ with entries
$b_{ij}$ of $B$ for $i\in I$ and $j\in J$.
Similarly, let $K=\br(\hat y)$ and $L=\supp(\hat y)$, and let
$A_{KL}$ be the corresponding $|K|\times|L|$ submatrix of~$A$.
Then the columns of $B_{IJ}$ are linearly independent, and 
the rows of $A_{KL}$ are linearly independent.
\abs{\rm(g)}
Consider any $\hat x\in X$ and $\hat y\in Y$.
Let $I$ be the set of labels of $\hat x$,
let $J$ be the set of labels of $\hat y$,
and let
\begin{equation}
\label{IJ}
\begin{array}{rcll}
P(I)&=&\{\,x\in P&\mid x \hbox{ has at least all the labels in }I\,\}\,,
\\
Q(J)&=&\{\,y\in Q&\mid y \hbox{ has at least all the labels in }J\,\}\,.
\end{array}
\end{equation}
Then $P(I)$ has dimension $m-|I|$, and $Q(J)$ has dimension
$n-|J|$.
\abs{\rm(h)}
$P$ and $Q$ are simple polytopes, and for both polytopes
any inequality that is redundant (that is, can be omitted
without changing the polytope) is never binding. 
\abs{\rm(i)}
$P$ and $Q$ are simple polytopes, and any pure strategy of a
player that is weakly dominated by or payoff equivalent to a
different mixed strategy is strictly dominated. 
\end{theorem}

\proof 
We show the implication chain
(a) $\Rightarrow$ (b), \ldots,
(h) $\Rightarrow$ (i),
(i) $\Rightarrow$ (a).

Assume (a), and consider any $y\in Q$.
If $Ay<\1$ then the only labels of $y$ are $m+j$ for
$y_j=0$ where $1\le j\le n$.
Hence, we can assume that at least one inequality of $Ay\le\1$
is binding, which corresponds to a best response (and hence
label) of the mixed strategy $\overline y=y\frac 1{\1\T y}$.
Via the projective map (\ref{project}), $\overline y$
and $y$ have the same labels.
By Proposition \ref{p-nondeg-label}, $\overline y$ 
and therefore $y$ has no more than $n$ labels, as claimed.
Similarly, no $x\in P$ has no more than $m$ labels.
This shows~(b). 

Assume (b); we show (c).
For the game $(C,C\T)$, the polytopes corresponding to
(\ref{defPQ}) are
\begin{equation}
\label{CPQ}
\arraycolsep0pt
\begin{array}{rllrlrl}
P'&{}=\{\,x'\in \reals^d&{}\mid {}&x'&{}\ge\0,~& Cx'&{}\le \1\},\\
Q'&{}=\{\,y'\in \reals^d&{}\mid {}&Cy'&{}\le\1,~&y'&{}\ge\0\}\,,
\end{array}
\end{equation}
so $P'=Q'=S$.
By (\ref{C}), for $z=(x,y)$ we have $z\ge\0$ and $Cz\le\1$
if and only if
\begin{equation}
\label{xyAB}
x\ge\0,
\quad 
y\ge\0,
\quad 
Ay\le\1,
\quad
B\T x\le\1,
\end{equation}
that is, $x\in P$ and $y\in Q$.
By (b), $x$ has no more than $m$ labels and $y$ has no more
than $n$ labels, so $z=(x,y)$ has no more than $m+n=d$
labels, and this holds correspondingly for any $x'\in P'$
and $y'\in Q'$ in (\ref{CPQ}).
Therefore, $(C,C\T)$ is nondegenerate.

Assume (c).
The inequalities of the polytope $P'$ in (\ref{CPQ}) have
unique labels $1,\ldots,2d$ (unlike $S$).
No point in $P'$ has more than $d$ labels, and therefore no
point in $S$ has more than $d$ binding inequalities.
This shows (d).

Assume (d), and, to show (e), suppose for some $z\in S$
with $K=\{\,i\mid z_i=0\}$ and $L=\{\,j\mid c_j\T z=1\}$
the row vectors
$e_i\T$ for $i\in K$ and $c_j\T$ for $j\in L$
are linearly dependent;
choose $z$ so that $|K|+|L|$ is maximal.
By (d), $|K|+|L|\le d$.
Let $U$ be the matrix with rows 
$e_i\T$ for $i\in K$ and $c_j\T$ for $j\in L$,
which has row rank $r<|K|+|L|\le d$,
and therefore only $r<d$ linearly independent columns.
Hence, there is some nonzero $v\in\reals^d$ so that
$Uv=\0$.
For $\alpha\in\reals$ let $z_\alpha=z+v\alpha$.
Then $(z_\alpha)_i=0$ for $i\in K$ and 
$c_j\T z_\alpha=1$ for $j\in L$ because $Uv=\0$.
For $\alpha=0$,
the inequalities $(z_\alpha)_i\ge 0$ for $i\not\in K$ and 
$c_j\T z_\alpha\le1$ for $j\not\in L$ are not binding,
but maximizing $\alpha$ subject to these inequalities (which
imply $z_\alpha\in S$) produces at least one further binding
inequality because $S$ is bounded and $v\ne\0$.
This contradicts the maximality of $|K|+|L|$. 
This proves (e).

Assume (e), and consider $\hat x,\hat y,I,J,K,L$ as defined
in~(f), with best-response payoff $u$ to player~1 and $v$ to
player~2.
Let $x=\hat x \frac1{v}$
and $y=\hat y \frac1{u}$ so that
$x\in P$ and $y\in Q$ via (\ref{project1}) and (\ref{project}).
With $z=(x,y)$, the binding inequalities in $z\ge\0$ and
$Cz\le\1$, that is, (\ref{xyAB}), are
$x_i=0$ for $i\not\in I$ and
$y_j=0$ for $j\not\in L$ and
$a_i\T y=1$ for $i\in K$ where $A\T=[a_1\cdots a_m]$ and
$b_j\T x=1$ for $j\in J$ where $B=[b_1\cdots b_n]$.
The corresponding row vectors $e_i\T$ for $i\not\in I\cup L$
and (as rows of $C$) $(\0,a_i\T)$ for $i\in K$ and
$(b_j\T,\0)$ for $j\in J$ are linearly independent by
assumption~(e).
This implies that the rows $a_{iL}\T$ of $A_{KL}$ are
linearly independent\,: suppose 
$\sum_{i\in K}\alpha_i\,a_{iL}\T=\0\T$ for some reals
$\alpha_i$.
Then with $\beta_j=-\sum_{i\in K}\alpha_i\,a_{ij}$ for $j\not\in L$
we have
$\sum_{j\not\in L}\beta_j\,e_j\T
+\sum_{i\in K}\alpha_i\,a_{i}\T=\0\T$
which by linear independence of these row vectors
is only the trivial linear combination, so $\alpha_i=0$ for
$i\in K$ as claimed.
Similarly, the columns of $(B_{IJ})\T$, that is, rows of
$B_{IJ}$, are linearly independent, as claimed in~(f).

Assume (f), and consider $\hat x,\hat y,I,J$ as in (g).
With the set $J$ of labels of $\hat y$, let
\begin{equation}
\label{KJL}
\begin{array}{rcl}
K&=&J\cap \{1,\ldots,m\},
\\
J'&=&\{\,j\in\{1,\ldots,n\}\mid m+j\in J\},
\\
L&=&\{\,j\in\{1,\ldots,n\}\mid m+j\not\in J\}, 
\end{array}
\end{equation}
that is, $K=\br(\hat y)$ and $L=\supp(\hat y)$, and
\begin{equation}
\label{J}
|J| =|K|+|J'|=|K|+n-|L|.
\end{equation}
Let $A_{KL}$ be the submatrix of $A$ with entries
$a_{ij}$ of $A$ for $i\in K$ and $j\in L$.
We write $y\in Q$ as $y=(y_{J'},y_L)$.
Then
\begin{equation}
\label{QJ}
y=(y_{J'},y_L)\in Q(J)
\quad\Leftrightarrow\quad
y_{J'}=\0,
\quad
y_L\ge\0, 
\quad
A_{KL}\,y_L=\1\,.
\end{equation}
The $|K|$ equations $A_{KL}\,y_L=\1$ with $|L|$ variables
are underdetermined, where we show that its solution set for
all constraints in (\ref{QJ}) has dimension $|L|-|K|$.
By assumption (f), $A_{KL}$ has full row rank $|K|$,
so there is an invertible $|K|\times|K|$ submatrix
$A_{KK'}$ of $A_{KL}$,
where we write
$A_{KL}=[A_{KK'}~A_{KL'}]$
and
$y_L=(y_{K'},y_{L'})$,
so that the following are equivalent:
\begin{equation}
\label{AKK}
\begin{array}{rcl}
A_{KL}\,y_L
&=&\1\,,
\\
A_{KK'}\,y_{K'}
+A_{KL'}\,y_{L'}
&=&\1\,,
\\
y_{K'}
&=&A_{KK'}^{-1}\1-(A_{KK'}^{-1}A_{KL'})y_{L'}~,
\end{array}
\end{equation}
where $y_{L'}$ can be freely chosen subject to 
$y_L=(y_{K'},y_{L'})\ge\0$ to ensure (\ref{QJ}).
Let
$\ell=|L'|=|L|-|K'|=|L|-|K|=n-|J|$ by (\ref{J}).
We claim that (\ref{QJ}) and (\ref{AKK}) imply that $Q(J)$
is a set of affine dimension~$\ell$.
By definition, this means that $Q(J)$ has $\ell+1$ (but no
more) points $y^0,y^1,\ldots,y^\ell$ that are affinely
independent, or equivalently (as is easy to see) that the
$\ell$ points 
\begin{equation}
\label{li}
y^1-y^0,\ldots,
y^\ell-y^0\quad
\hbox{are linearly independent.} 
\end{equation}
Any $y^j\in Q(J)$ is by (\ref{QJ}) of the form
$y^j=(y_{J'}^j\,,y_{K'}^j\,,y_{L'}^j)$,
where $y_{J'}^j=\0$ and $y_{K'}^j$ is an affine function of
$y_{L'}^j$ by (\ref{AKK}).
Hence, $y_{K'}^j-y_{K'}^0$ is a linear function of
$y_{L'}^j\in\reals^\ell$, and there can be no more than 
$\ell$ linearly independent vectors $y^j-y^0$ in (\ref{li}).
We find such vectors as follows.
Let $u$ be the best-response payoff to $\hat y$ and
$y^0=\hat y\frac1u$,
and (assuming for simplicity that $L'=\{1,\ldots,\ell\})$
$y^j=y^0+e_j\,\eps$ for $j\in L'$ and $\eps>0$.
Then $y^0\in Q(J)$ and for sufficiently small $\eps$
\begin{equation}
\label{strict}
y^j_{L}=(y^j_{K'}\,,y^j_{L'})>\0\,,
\qquad
a_i\T y^j<1 \quad\hbox{for }i\not\in K
\end{equation}
because these strict inequalities hold 
(as ``non-labels'' of $\hat y$) for $j=0$
and $y^j$ is by (\ref{AKK}) a continuous function of its
part $y^j_{L'}$ whose $j$th component is augmented
by~$\eps$.
Then $y^j-y^0$ are the scaled unit vectors $e_j\,\eps$ which
are linearly independent, which implies (\ref{li}).
So $Q(J)$ has dimension $\ell=n-|J|$.
Similarly, $P(I)$ has dimension $m-|I|$.
This shows (g).

Assume (g).
If, say, $Q$ was not simple, then some point $y$ of $Q$
would be on more than $n$ facets and have a set $J$ of more
than $n$ labels.
The corresponding set $Q(J)$ would have negative dimension
and be the empty set, but contains~$y$, a contradiction.
So $Q$, and similarly $P$, is a simple polytope.
Suppose some inequality of $Q$ is redundant, and that it is
sometimes binding, with label~$k$.
This binding inequality therefore defines a nonempty face
$F$ of~$Q$.
Consider the set $J$ of labels that \textit{all} points in
$F$ have, which includes~$k$.
Because the inequality is redundant, $Q(J)$ and $Q(J-\{k\})$
are the same set, but have different dimension by (g), a
contradiction.
The same applies to~$P$.
This shows (h).

Assume (h).
We show that because $Q$ has no redundant inequality that is
binding, player~1 has no pure strategy $i$ that is weakly
dominated by or payoff equivalent to a different mixed
strategy~$x\in X$, and not strictly dominated.
Suppose this was the case, that is,
\begin{equation}
\label{weakd}
a_i\T\le x\T A,
\qquad 
x_i=0
\end{equation}
where $a_i\T$ is the $i$th row of~$A$.
In (\ref{weakd}) we can assume $x_i=0$ by replacing, if
necessary, $x$ with $x-e_i x_i$ and re-scaling because
$x\ne e_i$.
Then the $i$th inequality $a_i\T y\le 1$ in $Ay\le\1$ is
redundant, because it is implied by the other inequalities
in $Ay\le\1$ since $y\ge\0$ implies
$a_i\T y\le x\T Ay\le x\T\1=1$.
Because $i$ is not strictly dominated by some mixed
strategy,
it is not hard to show (see Lemma~3 of \citealp{Pearce1984})
that $i$ is the best response to some mixed strategy
$\hat y\in Y$, with best response payoff $u$,
so $a_i\T \hat y=u$.
But then for $y=\hat y\frac1u\in Y$ the inequality
$a_i\T y\le 1$ is binding, which contradicts (h).
The same applies for $P$ and player~2.
This shows~(i). 

Finally, (i) implies (a) where we use
Proposition~\ref{p-nondeg-label}.
Any $\hat y$ in $Y$ with more than $n$ labels would, via
(\ref{project}), either define a point $y$ in $Q$ that is on
more than $n$ facets so that $Q$ is not simple, or one of
the labels would define the exact same facet as another
and thus a duplicate pure strategy, or 
one of the labels $i$ would define a lower-dimensional face
$F=\{y\in Q\mid a_i\T y = 1\}$ as in the implication
(g)~$\Rightarrow$~(h) which can be shown to imply
(\ref{weakd}) for some $x\in X$, all contradicting~(i).
The same applies for the other player. 
\endproof

In Theorem~\ref{t-eq}, condition (b) is very similar to
Proposition~\ref{p-nondeg-label}, but applies to the labels
of points in $P$ and $Q$ rather $X$ and $Y$.
Condition (f) (and similarly (e)) states full row rank of
the best-response submatrix $A_{KL}$ of the payoff
matrix~$A$ to player~1 for the support $L$ and best-response
set $K$ of a mixed strategy ${\hat y\in Y}$, and similarly for
the other player.
This uses the condition that $P$ and $Q$ are polytopes,
namely positive best-response payoffs by
Proposition~\ref{p-posbr}.
Otherwise, a nondegenerate game may have a payoff
(sub)matrix that does not have full rank, such as $\left[
\begin{matrix} -1 & \phantom-1\\ \phantom-1 & -1
\end{matrix} \,\right]$.

Condition (g) is about the dimension of the sets $P(I)$ and
$Q(J)$ defined by sets of labels $I$ and $J$.
These are the labels of some mixed strategies, which
ensures that $P(I)$ and $Q(J)$ are not empty.
The condition states that each extra label reduces the
dimension by one.
A singleton label set defines a facet of $P$ or~$Q$.
Condition (h) is also geometric, and is about the shape of
the polytope (being simple) and about its description by
linear inequalities.
For example, a duplicate strategy of player~1 and thus
duplicate row of $A$ would not change the shape of~$Q$, but
affect its labels.
Redundant inequalities are allowed as long as they do not
define labels at all.
In (i) these never-binding inequalities are strictly dominated
strategies.
Condition (\ref{weakd}) states that the pure strategy~$i$
of player~1 is weakly dominated by a different mixed
strategy~$x$, or payoff equivalent to it if ${a_i=x\T A}$.

\section{Pivoting and handling degenerate games}
\label{s-pivot}

As mentioned before the start of Section~\ref{s-LHpoly}, the
LH algorithm is a path-following method that can be
implemented by certain algebraic operations.
These are known as ``pivoting'' as used in the simplex
algorithm for linear programming (see \citealp{Dantzig1963},
or, for example, \citealp{MG}).
We explain this using the letters $A,B,m,n$ that are
standard in this context and do \textit{not} refer to a
bimatrix game.

Let $C$ be an $m\times d$ matrix and $q\in\reals^m$, and
consider,
like in (\ref{S}) (where we have assumed $m=d$)
the polyhedron
$S=\{z\in\reals^d\mid z\ge\0,~Cz\le q\,\}$.
Then $z\in S$ if and only if there is some $w\in\reals^m$ so
that 
\begin{equation}
\label{eqpoly}
Iw+Cz=q,
\qquad
w\ge\0,\qquad z\ge\0,
\end{equation}
where $I$ is the $m\times m$ identity matrix.
We write this more generally with $n=m+d$ and the $m\times n$ matrix
$A=[I~~C]$ and $x=(w,z)\in\reals^{n}$ as
\begin{equation}
\label{Axq}
Ax=q,\qquad x\ge\0. 
\end{equation}
Any $x\in\reals^n$ that fulfills (\ref{Axq}) is called
\textit{feasible} for these constraints.
A \textit{linear program} (LP) is the problem of maximizing 
a linear function $c\T x$ subject to (\ref{Axq}),
for some $c\in\reals^n$.

Let $A=[A_1\cdots A_n]$.
For any partition $B,N$ of $\{1,\ldots,n\}$ we write
$A=[A_B~A_N]$, $x=(x_B,x_N)$,
$c\T x=(c_B\T x_B, c_N\T\,x_N)$.
We say $B$ is a \textit{basis} of $A$ if $A_B$ is an
invertible $m\times m$ matrix
(which implies $|B|=m$ and $|N|=n-m=d$;
this requires that $A$ has full row rank, e.g.\ if $I$ is
part of $A$ as above).
Then the following equations are equivalent for any $x\in\reals^n$:
\begin{equation}
\label{bfs}
\arraycolsep.2em
\begin{array}{rcl}
Ax&=&q\,,
\\
A_B\,x_B+A_N\,x_N&=&q\,,
\\
x_B&=&A_B^{-1}q-A_B^{-1}A_N\,x_N\,.
\\
\end{array}
\end{equation}
For the given basis $B$ and $x=(x_B,x_N)$, the last equation
expresses how the ``basic variables'' $x_B$ depend on the
``nonbasic variables'' $x_N$ (so that $Ax=q$).
The \textit{basic solution} associated with $B$ is given by
$x_N=\0$ and thus $x_B=A_B^{-1}q$.
It is called \textit{feasible} if $x_B\ge\0$.

Basic feasible solutions are the algebraic representations
of the \textit{vertices} of the polyhedron defined
by~(\ref{Axq}), called $H$ in the following proposition;
for the system (\ref{eqpoly}) there is a bijection between $S$
and $H$ via $z\in S$ and $x=(w,z)\in H$ with $w=q-Cz$.

\begin{proposition}
Let $H=\{x \mid Ax=q,~x\ge\0\}$
be a polyhedron where $A$ has full row rank.
Then $x$ is a vertex of $H$ if and only if $x$ is a basic
feasible solution to~$(\ref{Axq})$.
\end{proposition}

\proof
Let $x=(x_B,x_N)$ be a basic feasible solution with basis
$B$ and consider the LP $\max_{x\in H} c\T x$ for $c_B=\0$,
$c_N=-\1$.
Then clearly $c\T x\le0$ for all $x\in H$ (so this is a
valid inequality for $H$) and $c\T x=0$ for the basic
solution, which is therefore optimal.
Moreover, $c\T x=0$ implies $c_N\T \,x_N=0$, which means the
only optimal solution is the basic solution
$(x_B,x_N)=(A_B^{-1}q,\0)$.
Hence the face $\{x\in H\mid c\T x=0\}$ has only one point
in it and is therefore a vertex.
This shows every basic feasible solution is a vertex.

Conversely, suppose $\hat x$ is a vertex of $H$, that is,
$\{x\in H\mid c\T x=q_0\}=\{\,\hat x\,\}$ where
$c\T x\le q_0$ is valid for $H$. 
Hence $\hat x$ is the unique optimal solution to the LP
$\max_{x\in H} c\T x$.
If the LP has an optimal solution then it has a basic
optimal solution (this can shown similarly to the
implication (d)~$\Rightarrow$~(e) for Theorem~\ref{t-eq}),
which equals~$\hat x$.
\endproof

In general, a vertex may correspond to several bases that
represent the same basic feasible solution, namely when 
at least one basic variable is zero and can be replaced by
some nonbasic variable. 
However, in a nondegenerate polyhedron the basis that
represents a vertex is unique.
This happens if and only if in any basic feasible solution
$(x_B,x_N)$ to (\ref{Axq}) with $x_N=\0$ we have $x_B>\0$,
which can be shown to be equivalent to
Theorem~\ref{t-eq}(e), for example, for the system
(\ref{eqpoly}).
For the moment, we assume this nondegeneracy condition.

For a vertex $z$ of the polytope $S$ in (\ref{S}), which
corresponds to $x=(w,z)\in H$ with a basic feasible solution
$x=(x_B,x_N)$, the \textit{binding inequalities} of
$z\ge\0$, $Cz\le\1$ correspond to the \textit{nonbasic
variables} $x_N=\0$ (because $x_B>\0$); these are exactly
$d=|N|$ binding inequalities.
We re-write (\ref{bfs}) as
\begin{equation}
\label{pivot}
x_B
~=~A_B^{-1}q-A_B^{-1}A_N\,x_N
~=~A_B^{-1}q-\sum_{j\in N}A_B^{-1}A_j\,x_j
~=:~\overline q-\sum_{j\in N}\overline A_j\,x_j 
\end{equation}
where $\overline q$ and $\overline A_j$ depend on the
basis~$B$.
In the basic feasible solution, $x_N=\0$.
In the LH algorithm as described in Section~\ref{s-LHpoly}, 
the next vertex is found by allowing one of the binding
inequalities to become non-binding.
This means that in (\ref{pivot}), one of the nonbasic
variables $x_j$ for $j\in N$ is allowed to increase from
zero to become positive.
This variable is called the \textit{entering} variable
(about to ``enter the basis''); all other nonbasic variables
stay zero.
The current basic variables $x_B$ then change linearly as
function of $x_j$ according to the equation and constraint
\begin{equation}
\label{enter}
x_B
~=~\overline q-\overline A_j\,x_j ~\ge~\0\,.
\end{equation}
When in this equation $x_j>0$ and $x_B>\0$, only $d-1$
inequalities are binding, which define an edge of~$S$.
Normally, for example if $S$ and thus $H$ is a polytope,
this edge ends at another vertex which is obtained when one
of the components $x_i$ of $x_B$ in (\ref{enter}) becomes
zero when increasing $x_j$.
Then $x_i$ is called the variable that \textit{leaves the
basis}, and the \textit{pivot step} is to replace $B$ with
$B\cup\{j\}-\{i\}$ which becomes the new basis which defines
the new vertex.
If the leaving variable $x_i$ is not unique, then at least
one other basic variable $x_k$ becomes simultaneously zero
with $x_i$, and is then a zero basic variable in the next
basis, which means a degeneracy.
Hence, for a nondegenerate polyhedron the leaving variable
is unique.

The pivot step is an algebraic representation of the edge
traversal.
In (\ref{enter}), the leaving variable is determined by
the constraints $\overline q_i-\overline a_{ij}\,x_j\ge0$
for the components $\overline q_i$ of $\overline q$ and
$\overline a_{ij}$ of $\overline A_j$, for $i\in B$.
These impose a restriction on $x_j$ only if $\overline
a_{ij}>0$ (if $\overline A_j\le\0$ then $x_j$ in
(\ref{enter}) can increase indefinitely, which would mean
that $H$ is unbounded, which we assume is not the case).
Hence, these constraints are equivalent to
\begin{equation}
\label{minratio}
\frac{\overline q_i}{\overline a_{ij}}\ge x_j\ge0,
\qquad \overline a_{ij}>0, \qquad i\in B\,.
\end{equation}
The smallest of the ratios $\overline q_i/\overline a_{ij}$
in (\ref{minratio}) thus determines how much $x_j$ can
increase to maintain the condition $x_B\ge\0$ in
(\ref{enter}).
Finding that minimum is called the \textit{minimum ratio
test}.
Moreover, that ratio is positive because the current basic
feasible solution is given by $x_B=\overline q>\0$.
The ratios in (\ref{minratio}) have a unique minimum which
determines the leaving variable.

Pivoting, the successive change from one basic feasible
solution to another by exchanging one ``entering'' nonbasic
variable for a unique ``leaving'' basic variable, thus
represents a path of edges of the given polytope~$S$.
In the LH algorithm, the entering variable is chosen
according to the following rule.

\begin{algorithm}[Lemke-Howson with complementary pivoting]
\label{a-LH}
Consider 
the system $(\ref{eqpoly})$ with $q=\1$ as in $(\ref{sym})$.
\einr 1.3em
\abs{1.}
Start with the basic feasible solution where $z=\0$, $w=q$.
Choose one $k$ as missing label which determines the first
entering variable~$z_k$.
\abs{2.}
In the pivot step, if the leaving variable is $w_k$ or
$z_k$, output the current basic solution and stop.
Otherwise, the leaving variable is $w_i$ or $z_i$ for
$i\ne k$.
Choose the complement of that variable ($z_i$ respectively
$w_i$) as the new entering variable and repeat step~2.
\end{algorithm}

This is the algebraic implementation of the LH algorithm.
It ensures that for each $i\ne k$ at least one variable
$w_i$ or $z_i$ is always nonbasic and represents a binding
inequality, so that the traversed vertices and edges of~$S$
have all labels except possibly~$k$.
Except for the endpoints of the computed path, both $w_k$
and $z_k$ are basic variables, which are positive throughout
and correspond to the missing label.

Pivoting has originally been invented by \citet{Dantzig1963}
for the simplex algorithm for solving an LP, where the
entering variable is chosen so as to improve the current
value of the objective function.
This is given as $c\T x=c_B\T x_B+c_N\T\,x_N$, and by
expressing $x_B$ as a function of $x_N$ according to
(\ref{bfs}), any $x_j$ with a positive coefficient can serve
as entering variable.
The optimum of the LP is found when there is no such
positive coefficient.
Hence, the only difference between the LH and the simplex
algorithm is the choice of the entering variable by the
``complementarity rule'' in step~2 above.

The LH algorithm, like the simplex algorithm, can be
generalized to the degenerate case where basic feasible
solutions may have zero basic variables.
For that purpose, the right-hand side $q$ in (\ref{Axq}) is
\textit{perturbed} by replacing it by
$q+(\eps,\eps^2,\ldots,\eps^m)\T$ for some sufficiently
small $\eps>0$ (which in the end can be thought of as
``vanishingly small'').
For a basis $B$, the corresponding basic solution
$(x_B,x_N)$ is given by $x_N=\0$ and
\begin{equation}
\label{lex}
x_B=A_B^{-1}q+A_B^{-1}(\eps,\eps^2,\ldots,\eps^m)\T
\end{equation}
and it is feasible (that is, $x_B\ge\0$) if and only if the
$m\times (1+m)$ matrix
\begin{equation}
\label{lexpos}
\hbox{
$[A_B^{-1}q~~A_B^{-1}]$ is \textit{lexico-positive}},
\end{equation}
that is, the first nonzero entry of each row of this matrix
is positive.
Note that $\overline q=A_B^{-1}q$ may have zero entries, but
$A_B^{-1}$ cannot have an all-zero row, so (\ref{lexpos})
implies $x_B>\0$ for all sufficiently small positive~$\eps$
in (\ref{lex}), and thus \textit{nondegeneracy throughout}.
However, no actual perturbance is needed, because
(\ref{lexpos}) is recognized solely from $A_B^{-1}$.
Condition (\ref{lexpos}) is maintained by extending
(\ref{minratio}) to a ``lexico-minimum ratio test'', which
determines the leaving variable uniquely
\citep[p.~1741]{vS2002}.
In that way, the LH algorithm proceeds uniquely even for a
degenerate game, and terminates at a Nash equilibrium.

For an accurate computation of the LH steps, it is necessary
to store the system (\ref{pivot}) precisely without rounding
errors as they may occur in floating-point arithmetic.
If the entries of the given bimatrix game are integers, then it is
possible to store this linear system using only integers and
a separate integer for the determinant of the current
basis matrix $A_B$.
This ``integer pivoting'' \citep[see][Section 3.5]{vS2007}
avoids numerical errors by storing arbitrary-precision
integers without the costly cancellation operations when
adding fractions in rational arithmetic.

Complementary pivoting as described in Algorithm~\ref{a-LH}
has been generalized by \citet{Lemke1965} to solve 
linear complementary problems (\ref{LCP}) for more general
parameters $C$ and~$q$.
The system (\ref{eqpoly}) is thereby extended by an
additional matrix column and variable~$z_0$.
A first basic solution has $w=q$ and $z=\0$ and $z_0$
which fulfills the complementarity condition $z\T w=0$
but is not feasible if $q$ has negative components.
Then $z_0$ enters the basis so has to obtain feasibility,
with some $w_i$ as leaving variable.
Then as in step~2 of Algorithm~\ref{a-LH}, the next entering
variable is $z_i$, more generally the complement of the
leaving variable, which is repeated until $z_0$ leaves the
basis.
A number of conditions on $C$ can ensure that there is no
``ray termination'', that is, the ``entering column''
$\overline A_j$ in (\ref{enter}) has always at least one
positive component (see \citealp{CPS}).

Most path-following methods that find an equilibrium of a
two-player game can be encoded as special cases of Lemke's
algorithm, such as \citet{GW2003}.
In \citet{vSET02} it is shown how to use it for mimicking the
(linear) ``Tracing Procedure'' of \citet{harsanyiselten}
that traces a path of best responses against a suitable
convex combination of a ``prior'' mixed-strategy pair as
starting point and the currently played strategies; it
terminates when the weight of the prior (encoded by the
variable~$z_0$) becomes zero.
Moreover, this algorithm can also be applied to more general
strategy sets, such as the ``sequence form'' for extensive
form games \citep{vs1996}.

\section{Maximal Nash subsets and finding all equilibria}
\label{s-maxnash}

The LH algorithm finds (at least) one Nash equilibrium of a
bimatrix game.
\textit{All} equilibria are found by
Algorithm~\ref{a-suppenum}, which checks the possible
support sets of an equilibrium.
This can be improved by considering instead of these support
sets the vertices of the labeled polytopes $P$ and $Q$ in
(\ref{defPQ}).

A degenerate bimatrix game may have infinite sets of Nash
equilibria.
They can be described via maximal \textit{Nash subsets}
\citep{Millham1974,winkels,Jansen1981}, called ``sub-solutions'' by
\citet{Nash1951}.
A Nash subset for $(A,B)$ is a nonempty product set $S\times
T$ where $S\subseteq X$ and $T\subseteq Y$ so that every
$(x,y)$ in $S\times T$ is an equilibrium of $(A,B)$; in
other words, any two equilibrium strategies $x\in S$ and
$y\in T$ are ``exchangeable''.
The following proposition shows that a maximal Nash subset
is just a pair of faces of $P$ and $Q$ that together have
all labels $1,\ldots,m+n$.

\begin{proposition}
\label{p-faces}
Let $(A,B)$ an $m\times n$ bimatrix game
with polytopes $P$ and $Q$ in
$(\ref{defPQ})$, and for $I,J\subseteq\{1,\ldots,m+n\}$ let
$P(I)$ and $Q(J)$ be defined as in $(\ref{IJ})$.
Then $(x,y)\in P\times Q-\{\0,\0\}$, re-scaled to a
mixed-strategy pair in $X\times Y$, is a Nash equilibrium if
and only if for some $I$ and~$J$ we have
\begin{equation}
\label{xyIJ}
(x,y)\in P(I)\times Q(J),
\qquad
I\cup J=\{1,\ldots,m+n\}\,.
\end{equation}
\end{proposition}

\proof
This follows from Proposition~\ref{p-completely}:
(\ref{xyIJ}) implies that $(x,y)$ is completely labeled and
therefore a Nash equilibrium.
Conversely, if $(x,y)$ is a Nash equilibrium and $I$ and $J$
are the set of labels of $x$ and $y$ (this may increase the
sets $I$ and~$J$ when starting from (\ref{xyIJ})), then
(\ref{xyIJ}) holds.
\endproof

In (\ref{xyIJ}), $P(I)$ is the face of $P$ defined by the
binding inequalities in $I$, and $Q(J)$ is the face of $Q$
defined by the binding inequalities in $J$.
In a nondegenerate game, these faces are vertices of $P$ and~$Q$.
In general, they may be higher-dimensional faces such as 
edges.
Usually, when the dimension of these faces is not too high,
it is informative to describe them via the vertices of these
faces, which are also vertices of $P$ or $Q$.
They are usually called \textit{extreme} equilibria.

\begin{proposition}[\citealp{winkels,Jansen1981}]
\label{p-ext}
Under the assumptions of Proposition~\ref{p-faces},
$(x,y)$ is, after re-scaling, a Nash equilibrium if and only
if there is a set $U$ of vertices of $P-\{\0\}$ and 
a set $V$ of vertices of $Q-\{\0\}$ so that 
$x\in\conv U$ and $y\in\conv V$, and every $(u,v)\in U\times
V$ is completely labeled.
\end{proposition}

\proof
$U$ and $V$ are just the vertices of $P(I)$ and $Q(J)$ in
(\ref{xyIJ}); see \citet[Prop.~4]{ARSvS}.
\endproof

Proposition~\ref{p-ext} shows that the set of all Nash
equilibria is completely described by the (finitely
many) extreme Nash equilibria.
Consider the bipartite graph $R$ on the vertices of
$P-\{\0\}$ and $Q-\{\0\}$ are the completely labeled vertex
pairs $(x,y)$, which are the extreme equilibria of $(A,B)$.
The maximal ``cliques'' (maximal complete bipartite subgraphs)
of $R$ of the form $U\times V$ then define the maximal Nash
subsets $\conv U\times\conv V$, as in
Proposition~\ref{p-ext}, whose union is the set of all Nash
equilibria. 
Maximal Nash subsets may intersect,
in which case their vertex sets intersect.
The inclusion-maximal connected sets of Nash equilibria are
the topological \textit{components}.
An algorithm that outputs the extreme Nash equilibria,
maximal Nash subsets, and components of a bimatrix game is
described in \citet{ARSvS} and available on the web at
{\small\url{http://banach.lse.ac.uk}} (at the time of this writing
for games of size up to $15\times 15$, due to the typically
exponential number of vertices that have to be checked).



\subsection*{Acknowledgment}
I thank Yannick Viossat for detailed comments.

\bibliographystyle{book} 
\small
\bibsep.75ex plus.1ex minus.05ex 
\bibliography{bib-eig}

\begin{thebibliography}{23}
\providecommand{\natexlab}[1]{#1}
\providecommand{\url}[1]{\texttt{#1}}
\providecommand{\urlprefix}{URL }

\bibitem[{Avis, Rosenberg, Savani, and von Stengel(2010)}]{ARSvS}
Avis, D., G.~D. Rosenberg, R.~Savani, and B.~von Stengel (2010).
\newblock Enumeration of {N}ash equilibria for two-player games.
\newblock \emph{Economic Theory} 42(1), 9--37.

\bibitem[{Cottle, Pang, and Stone(1992)}]{CPS}
Cottle, R.~W., J.-S. Pang, and R.~E. Stone (1992).
\newblock \emph{The Linear Complementarity Problem}.
\newblock Academic Press, Boston.

\bibitem[{Dantzig(1963)}]{Dantzig1963}
Dantzig, G.~B. (1963).
\newblock \emph{Linear Programming and Extensions}.
\newblock Princeton University Press, Princeton, NJ.

\bibitem[{Gale, Kuhn, and Tucker(1950)}]{GKT}
Gale, D., H.~W. Kuhn, and A.~W. Tucker (1950).
\newblock On symmetric games.
\newblock In: \emph{Contributions to the Theory of Games, Vol.~I}, edited by
  H.~W. Kuhn and A.~W. Tucker, volume~24 of \emph{Annals of Mathematics
  Studies}, 81--87. Princeton University Press, Princeton, NJ.

\bibitem[{Govindan and Wilson(2003)}]{GW2003}
Govindan, S. and R.~Wilson (2003).
\newblock A global {N}ewton method to compute {N}ash equilibria.
\newblock \emph{Journal of Economic Theory} 110(1), 65--86.

\bibitem[{Harsanyi and Selten(1988)}]{harsanyiselten}
Harsanyi, J.~C. and R.~Selten (1988).
\newblock \emph{A General Theory of Equilibrium Selection in Games}.
\newblock MIT Press, Cambridge MA.

\bibitem[{Jansen(1981)}]{Jansen1981}
Jansen, M. J.~M. (1981).
\newblock Maximal {N}ash subsets for bimatrix games.
\newblock \emph{Naval Research Logistics Quarterly} 28(1), 147--152.

\bibitem[{Keiding(1997)}]{Keiding1997}
Keiding, H. (1997).
\newblock On the maximal number of {N}ash equilibria in an $n\times n$ bimatrix
  game.
\newblock \emph{Games and Economic Behavior} 21(1-2), 148--160.

\bibitem[{Krohn, Moltzahn, Rosenm{\"u}ller, Sudh{\"o}lter, and
  Wallmeier(1991)}]{Krohn}
Krohn, I., S.~Moltzahn, J.~Rosenm{\"u}ller, P.~Sudh{\"o}lter, and H.-M.
  Wallmeier (1991).
\newblock Implementing the modified {LH} algorithm.
\newblock \emph{Applied Mathematics and Computation} 45(1), 31--72.

\bibitem[{Lemke(1965)}]{Lemke1965}
Lemke, C.~E. (1965).
\newblock Bimatrix equilibrium points and mathematical programming.
\newblock \emph{Management Science} 11(7), 681--689.

\bibitem[{Lemke and Howson(1964)}]{LH}
Lemke, C.~E. and J.~T. Howson, Jr (1964).
\newblock Equilibrium points of bimatrix games.
\newblock \emph{Journal of the Society for Industrial and Applied Mathematics}
  12(2), 413--423.

\bibitem[{{Matou\v sek} and G\"artner(2007)}]{MG}
{Matou\v sek}, J. and B.~G\"artner (2007).
\newblock \emph{Understanding and Using Linear Programming}.
\newblock Springer, Berlin.

\bibitem[{Millham(1974)}]{Millham1974}
Millham, C. (1974).
\newblock On {N}ash subsets of bimatrix games.
\newblock \emph{Naval Research Logistics Quarterly} 21(2), 307--317.

\bibitem[{Nash(1951)}]{Nash1951}
Nash, J.~F. (1951).
\newblock Non-cooperative games.
\newblock \emph{Annals of Mathematics} 54(2), 286--295.

\bibitem[{Pearce(1984)}]{Pearce1984}
Pearce, D.~G. (1984).
\newblock Rationalizable strategic behavior and the problem of perfection.
\newblock \emph{Econometrica} 52(4), 1029--1050.

\bibitem[{Shapley(1974)}]{Shapley1974}
Shapley, L.~S. (1974).
\newblock A note on the {L}emke-{H}owson algorithm.
\newblock \emph{Mathematical Programming Study 1: Pivoting and Extensions,}
  175--189.

\bibitem[{van Damme(1991)}]{vandamme-book2}
van Damme, E. (1991).
\newblock \emph{Stability and Perfection of {N}ash Equilibria}.
\newblock Springer-Verlag, Berlin, second edition.

\bibitem[{von Stengel(1996)}]{vs1996}
von Stengel, B. (1996).
\newblock Efficient computation of behavior strategies.
\newblock \emph{Games and Economic Behavior} 14(2), 220--246.

\bibitem[{von Stengel(2002)}]{vS2002}
von Stengel, B. (2002).
\newblock Computing equilibria for two-person games.
\newblock In: \emph{Handbook of Game Theory with Economic Applications}, edited
  by R.~J. Aumann and S.~Hart, volume~3, 1723--1759. North-Holland, Amsterdam.

\bibitem[{von Stengel(2007)}]{vS2007}
von Stengel, B. (2007).
\newblock Equilibrium computation for two-player games in strategic and
  extensive form.
\newblock In: \emph{Algorithmic Game Theory}, edited by N.~Nisan,
  T.~Roughgarden, E.~Tardos, and V.~Vazirani, 53--78. Cambridge University
  Press, Cambridge, UK.

\bibitem[{von Stengel(2021)}]{vS2021}
von Stengel, B. (2021).
\newblock \emph{Game Theory Basics}.
\newblock Cambridge University Press, Cambridge, UK.

\bibitem[{von Stengel, van~den Elzen, and Talman(2002)}]{vSET02}
von Stengel, B., A.~van~den Elzen, and D.~Talman (2002).
\newblock Computing normal form perfect equilibria for extensive two-person
  games.
\newblock \emph{Econometrica} 70(2), 693--715.

\bibitem[{Winkels(1979)}]{winkels}
Winkels, H.~M. (1979).
\newblock An algorithm to determine all equilibrium points of a bimatrix game.
\newblock In: \emph{Game Theory and Related Topics}, edited by O.~Moeschlin and
  D.~Pallaschke, 137--148. North-Holland, Amsterdam.

\end{thebibliography}

\end{document}